\documentclass[preprint,10pt,authoryear]{elsarticle}
\usepackage[left=1in, right=1in]{geometry}  

\usepackage{amsmath}
\usepackage{graphicx}
\usepackage{hyperref}
\usepackage{amssymb}
\usepackage{xcolor}
\usepackage{caption} 
\usepackage{algorithm} 
\usepackage{algpseudocode} 
\usepackage{adjustbox}
\usepackage{multirow}
\usepackage{array}
\newcommand{\tabularfont}{\footnotesize}
\usepackage{float} 
\usepackage{enumitem}

\usepackage{bm}
\usepackage{caption}
\usepackage{cleveref}
\crefname{figure}{Fig.}{Figs.}

\journal{Expert Systems with Applications}

\begin{document}

\begin{frontmatter}

\title{Physics-Constrained Graph Neural Networks for Spatio-Temporal Prediction of Drop Impact on OLED Display Panels}

\author[inst1]{Jiyong Kim}

\affiliation[inst1]{organization={Cho Chun Shik Graduate School of Mobility},
            addressline={Korea Advanced Institute of Science and Technology}, 
            city={Daejeon},
            postcode={34051}, 
            country={Republic of korea}}

\author[inst1]{Jangseop Park}

\author[inst2]{Nayong Kim}
\author[inst2]{Younyeol Yu}
\author[inst2]{Kiseok Chang}
\author[inst2]{Chang-Seung Woo}

\author[inst1]{Sunwoong Yang\corref{cor1}}
\author[inst1,inst3]{Namwoo Kang\corref{cor1}}

\cortext[cor1]{Co-corresponding authors}

\affiliation[inst2]{organization={LG Display},
            addressline={10, Magokjungang 10-ro}, 
            city={Seoul},
            postcode={07796}, 
            country={Republic of korea}}

\affiliation[inst3]{organization={Narnia Labs},
            addressline={193, Munji-ro}, 
            city={Daejeon},
            postcode={34051}, 
            country={Republic of korea}}

\begin{abstract}

This study aims to predict the spatio-temporal evolution of physical quantities observed in multi-layered display panels subjected to the drop impact of a ball. To model these complex interactions, graph neural networks have emerged as promising tools, effectively representing objects and their relationships as graph structures. In particular, MeshGraphNets (MGNs), state-of-the-art architectures for dynamic physics simulations using irregular mesh data, excel at capturing such dynamics. However, conventional MGNs often suffer from non-physical artifacts, such as the penetration of overlapping objects. To resolve this, we propose a physics-constrained MGN that mitigates these penetration issues while maintaining high level of accuracy in temporal predictions. Furthermore, to enhance the model's robustness, we explore noise injection strategies with varying magnitudes and different combinations of targeted components, such as the ball, the plate, or both. In addition, our analysis on model stability in spatio-temporal predictions reveals that during the inference, deriving next time-step node positions by predicting relative changes (e.g., displacement or velocity) between the current and future states yields superior accuracy compared to direct absolute position predictions. This approach consistently shows greater stability and reliability in determining subsequent node positions across various scenarios. Building on this validated model, we evaluate its generalization performance by examining its ability to extrapolate with respect to design variables. Furthermore, the physics-constrained MGN serves as a near real-time emulator for the design optimization of multi-layered OLED display panels, where thickness variables are optimized to minimize stress in the light-emitting materials. It outperforms conventional MGN in optimization tasks, demonstrating its effectiveness for practical design applications. In summary, the proposed physics-constrained MGN exhibits superior accuracy, robustness, and generalizability in drop impact scenarios compared to conventional MGN, demonstrating strong potential for real-time spatio-temporal prediction and design optimization.

\end{abstract}


\end{frontmatter}

\section{Introduction}
\label{sec:introduction}

Multi-layered display panels, commonly used in electronic devices such as smartphones, tablets, and televisions, require rigorous testing to ensure their durability. Among the experimental procedures for evaluating the mechanical properties of these panels, the dynamic ball drop test for determining structural damages is indispensable. This test is essential for improving impact resistance by minimizing stress distribution in light-emitting materials. However, this experimental approach is expensive and time-consuming, as it requires repeated trials and specialized instruments, such as high-speed cameras and sensors. While finite element method (FEM)-based computer simulations using explicit dynamics are widely used as an efficient alternative \citep{Wu1998, Pan2003, Hwan2011, Liu2018, Fan2021, Wu2022}, they are still limited in their ability to evaluate performance in real-time manner due to the high computational cost stemmed from the complexity and nonlinearity of the physics. These constraints highlight the need for the innovative methods to evaluate or optimize panel designs with real-time manner.

To address these challenges, data-driven surrogate modeling has been proposed as a promising solution. It allows for the rapid prediction of quantities in complex physical systems in real-time, thereby efficiently verifying product performance \citep{asher2015review, zhang2022deep, kang2022physics, shin2023wheel, yang2023inverse, he2024deep, lang2024physics,yang2022design}. Building on this approaches, \cite{Kim2024} employed machine learning-based surrogate models, including linear regression, CatBoost \citep{prokhorenkova2018catboost}, and random forest \citep{breiman2001random}, to predict the flexibility and durability of flexible AMOLED panels with varying thicknesses and material properties under bending and drop tests. \cite{Kim2024} demonstrated that these models could rapidly infer maximum principal strain in multi-layered display panels under ball drop impact, thereby accelerating the mechanical evaluation procedure for optimizing panel structures. Although these machine learning models are proficient in delivering quick predictions with high levels of accuracy, they are primarily designed to learn simple, static input-output relationships, which limits their ability to model the intricate spatio-temporal evolution of the quantities within the various mesh geometries. This limitation hinders their ability to predict continuous changes over time, such as tracking transient movement of the objects during a drop impact or capturing the localized deformation and stress patterns under dynamic loading conditions.

Similarly, \cite{Wadagbalkar2021} leveraged neural networks and decision tree regression models to predict projectile residual velocity in real-time. Their models utilized initial velocity and categorical features such as projectile shape, laminate orientation, and impact angle as input. Although these models incorporated dynamic characteristics by predicting residual velocity over time steps, they still limited the explicit dynamics simulation to scalar values, such as residual velocity, using scalar inputs, similar to \cite{Kim2024}'s approach. As a result, these models struggled to capture the complex spatial distribution of physical quantities across the entire structure during impact, particularly stress distribution and plastic deformation.

To overcome such pain points, researchers across various domains have been exploring data-driven surrogate models that enable spatio-temporal predictions \citep{tekin2021spatio, Wang2022, wang2024robust, wu2024spatio, zhao2024interpretable}. These models should be better suited for capturing both the spatial distribution and temporal evolution of physical quantities, providing more detailed and accurate predictions of dynamic responses and structural integrity in complex systems. In this context, \cite{Shao2023} developed two approaches to accelerate structural analysis from a spatio-temporal perspective: a fully connected neural network (FCNN) for predicting stress distribution across the target plate in the spatial domain, and a bidirectional LSTM (Bi-LSTM) network for predicting the stress-time sequence during bullet impact in the temporal domain. While these surrogate models demonstrated high accuracy in predicting stress, closely aligning with finite element analysis (FEA) results, the proposed model had constraints in fully capturing the dynamics.  The FCNN, for example, could only predict the stress distribution at the final time step for a fixed mesh, lacking the ability to generalize across different spatial configurations. Moreover, the Bi-LSTM, despite its temporal capabilities, could only predict stress for randomly selected elements on the target plate, rather than for the entire system. Consequently, these models faced challenges in fully capturing the dynamics and providing a comprehensive, mesh-agnostic spatio-temporal prediction.

Recently, graph neural networks (GNNs) are gaining significant attention since they are inherently capable of modeling the spatial relationships in complex, irregular geometries while also tracking the temporal evolution of physical quantities \citep{yang2024enhancing}. One such approach is MeshGraphNets (MGNs), a GNN-based deep learning model proposed by \cite{Pfaff2020}, which accurately predict the future states of irregular meshes, thereby enabling spatio-temporal predictions. This approach has been successfully applied in various mechanical fields such as computational fluid dynamics (CFD) \citep{gao2024finite, li2024learning, pegolotti2024learning}, weather forecasting \citep{lam2023learning}, cloth simulation \citep{grigorev2023hood, liao2024senc}, and solid mechanics \citep{wang2024towards, dalton2023physics}.

Despite the promising capabilities of MGNs, a significant issue can arise during ball drop impact simulations: the prediction of non-physical phenomena where flexible bodies (e.g., the ball and multi-layered display panels in our case) penetrate each other. This penetration accumulates errors over time, leading to degradation of long-term prediction accuracy and compromising the model's reliability for simulating impacts. To resolve this issue, we propose a physics-constrained MGN that mitigates such non-physical penetration between flexible bodies at each incremental time step during these simulations. First, since our proposed model is based on the conventional MGNs, it represents different mesh geometries as graphs and employs an encoder-processor-decoder (EPD) architecture to effectively capture the complex interactions between two flexible bodies and dynamic behavior of irregular meshes. The distinguishing feature of our approach is the integration of domain-specific constraints into the loss function. This is achieved by identifying critical boundary nodes, approximating boundary lines using polynomial fitting, and calculating the penetration depth between objects --- the degree of the penetration is then served as the additional loss function, which prevents the penetration of two objections. It is expected that this methodology will guarantee that the model adheres to the fundamental physical principles of the simulation, thereby effectively mitigating issues pertaining to object penetration.

To demonstrate the effectiveness of our proposed model, we first analyze the performance and inherent limitations in term of object penetration of the vanilla MGN, underscoring the need for a physics-constrained approach. A parametric study is then conducted to validate the accuracy of the physics-constrained approach in simulating the drop impact of a ball on a multi-layered display panel. This is followed by a detailed comparative evaluation between the vanilla MGN and the physics-constrained MGN, with emphasis on computational cost and predictive accuracy in complex drop impact scenarios. The robustness of our model is further showcased through a case study involving noise injection at varying magnitudes and components, aimed at mitigating error accumulation across temporal predictions. In addition, to enhance model stability and reliability in predicting subsequent node positions over time during the inference, we compare two approaches: predicting relative changes (displacement or velocity) between current and future states, and directly predicting absolute node positions. Finally, we evaluate the model's generalization capabilities by examining its performance in extrapolating beyond the initial range of design variables. To further demonstrate the model's practical utility, we couple the proposed predictive model with a gradient-free optimization algorithm, creating a real-time emulator to find optimal configurations of multi-layered display panels. Specifically, we optimize the design variables (thickness of the optically clear adhesive (OCA) layers between multi-layer display panels in our study) to minimize stress distribution within light-emitting materials, leveraging our pretrained predictive models. Our results demonstrate that the proposed model maintains higher accuracy in optimization outcomes compared to a vanilla model when validated against simulation data.

\clearpage
\noindent The main contributions of this study can be summarized as follows:

\begin{enumerate}
\item We develop a physics-constrained MGN model, which integrates domain-specific physical loss functions to mitigate non-physical penetration phenomena during drop impact tests between two flexible bodies.

\item A comprehensive parametric study is conducted to examine the effects of polynomial fitting for each component and the weights of the physics constraint loss---both integral requirements of the physics-constrained MGN---on its overall performance, providing valuable insights and guidelines for future researchers.

\item We further improve the performance of the physics-constrained MGN by determining the optimal noise magnitude to inject into the training dataset and applying this noise across all components involved in the drop impact test. This approach improves the model in terms of error accumulation, thereby ensuring its robust predictive accuracy over long-term predictions.

\item Our analysis on model stability in spatio-temporal predictions demonstrates that inferring future node positions by predicting relative changes (e.g., displacement or velocity) between current and future states consistently outperforms direct absolute position predictions.

\item To evaluate the model's generalization performance, we test the model beyond the training range with respect to the design variables, successfully verifying the reliable performance even in the extrapolated conditions.

\item We apply our proposed model to optimize the OCA thickness of multi-layered display panels, demonstrating its effectiveness in design optimization. By achieving more accurate predictions of maximum stress and deformation, the physics-constrained MGN again proves its superior performance over the vanilla MGN.

\end{enumerate}

\section{Physics-constrained MeshGraphNet}
\label{sec:methodology}

\begin{figure}[H]
    \centering
    \includegraphics[width=1\linewidth]{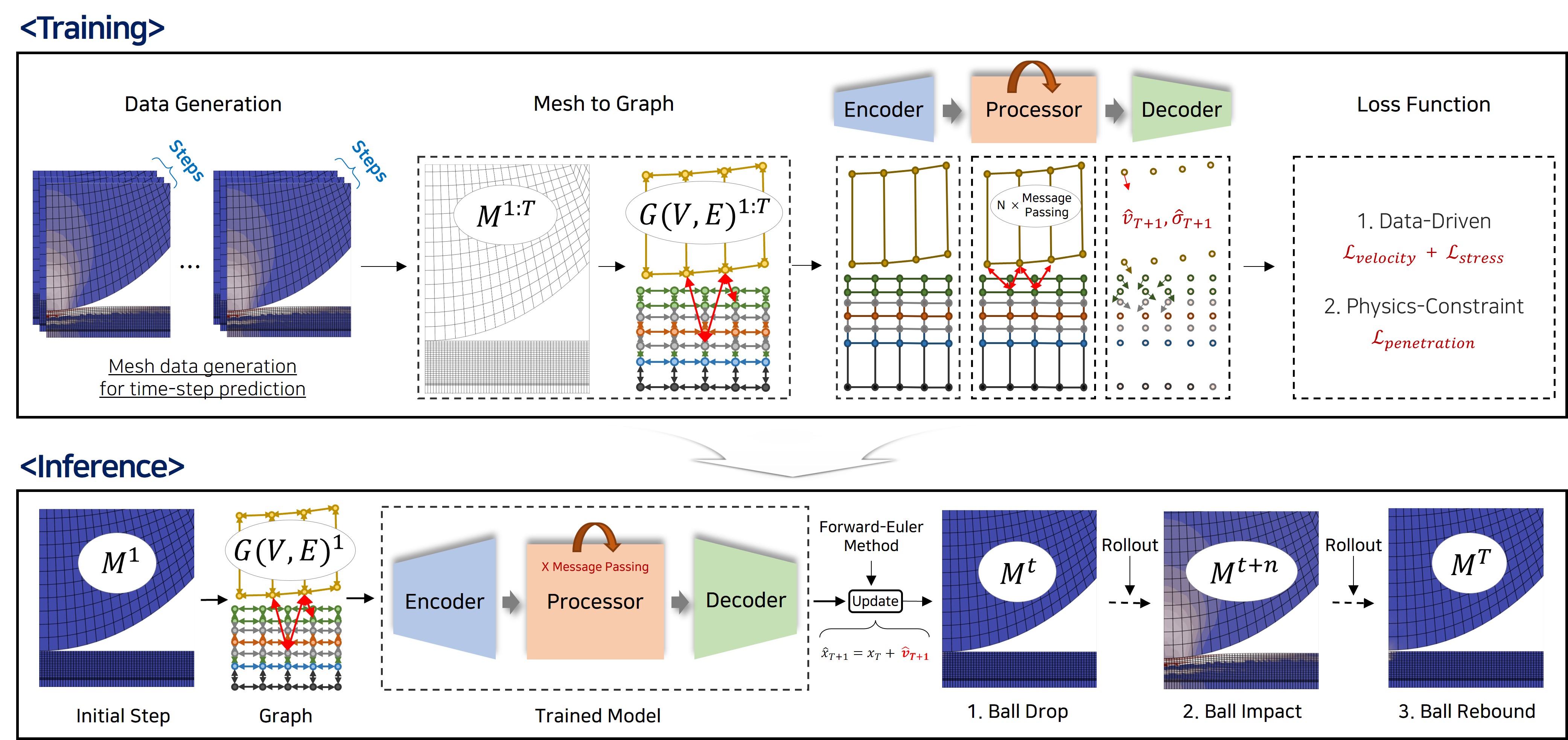}
    \captionsetup{justification=raggedright,singlelinecheck=false}
    \caption{Two-phase flowchart illustrating the training and inference processes of the physics-constrained MGN.}
    \label{fig:Fig.1}
\end{figure}

Our framework for predicting the nonlinear dynamics of multi-layered display panels under drop impact integrates graph-based learning with physics constraints. As depicted in \cref{fig:Fig.1}, the framework transforms mesh data \textit{M} into a graph \textit{G(V, E)}. This graph comprises a set of nodes \textit{V} and a set of edges \textit{E}, which represent the connections between adjacent nodes. The mesh-to-graph conversion is essential, as it simplifies the data structure, preserves spatial relationships, and enables efficient computation and model training with MGNs. Building on this graph structure, our model employs the encoder-processor-decoder (EPD) architecture proposed by \cite{sanchez2020learning}, facilitating effective learning and inference in drop impact simulations. In this study, we specifically tailor the predictive model to enhance performance through embedding a penetration physics constraint into the loss function. This constraint is included as a penalty term, guiding the model to produce outputs consistent with physical laws. As a result, integrating this physics constraint aims to improve the accuracy, stability, and generalization of the model’s predictions, ensuring they align with the underlying physics of the problem.

This section proceeds as follows: \Cref{sec:2.1} details the step-by-step procedure for converting the mesh data into a graph structure, highlighting the key considerations in this transformation. \Cref{sec:2.2} provides an in-depth analysis of the proposed encoder-processor-decoder architecture, emphasizing its components and their roles in capturing complex system dynamics. \Cref{sec:2.3} discusses the formulation of the physics-constrained loss function, with a focus on integrating physical constraints of penetration as penalty terms.

\subsection{Converting mesh to graph structure}
\label{sec:2.1}

The process of transforming the mesh \( M \) into a graph representation \( G(V, E^M, E^C) \) is outlined in Algorithm~\ref{alg:mesh_to_graph}. This conversion is essential for structuring the data in a format that is optimal for input into EPD architecture.

\begin{algorithm}[H]
\small
\caption{Conversion from mesh to graph structure}
\label{alg:mesh_to_graph}
    \renewcommand{\algorithmicrequire}{\textbf{Input:}}
    \renewcommand{\algorithmicensure}{\textbf{Output:}}
    
    \newcommand{\algorithmicnodefeatures}{\textbf{Node features:}}
    \newcommand{\NodeFeatures}{\item[\algorithmicnodefeatures]}
    \newcommand{\algorithmicedgefeatures}{\textbf{Edge features:}}
    \newcommand{\EdgeFeatures}{\item[\algorithmicedgefeatures]}

    \begin{algorithmic}[1]

    \Require Mesh \( M = (C_{\text{quad}}, x^m, x^w, n_i) \), contact radius \( r_w \)
    
    \Ensure Graph \( G(V, E^M, E^C) \), where \( V \) is the set of nodes \( v_i \), \( E^M \) is the set of mesh edges \( e_{ij}^m \), and \( E^C \) is the set of contact edges \( e_{ij}^c \)
    
    \NodeFeatures
    \State \( v_i \leftarrow \text{CONCAT}\left( \varphi(n_i), \ x_i^w, \ \dot{x}_i^w \right) \)
    \Comment{Create node features by combining node type, position, and velocity}
    
    \EdgeFeatures
    \State Extract edge pairs \( (i, j) \) from \( C_{\text{quad}} \)
    \Comment{Obtain edges from quadrilateral cells}

    \Statex \textbf{Mesh edge features:}
        \ForAll{mesh edges \( (i, j) \)}
            \State \( u_{ij} \leftarrow x_i^m - x_j^m \) \Comment{Relative position in mesh space}
            \State \( x_{ij}^w \leftarrow x_i^w - x_j^w \) \Comment{Relative position in world space}
            \State \( e_{ij}^m \leftarrow \text{CONCAT}(u_{ij}, \ |u_{ij}|, \ x_{ij}^w, \ |x_{ij}^w|) \) \Comment{Form mesh edge features}
            \State \( e_{ij}^m \leftarrow \text{NORMALIZE}(e_{ij}^m) \) \Comment{Normalize features}
        \EndFor
        
    \Statex \textbf{Contact edge features:}
        \State \( \text{Compute distances } D_{ij} = \| x_i^w - x_j^w \|_2^2 \)
        \State \( \text{Exclude node pairs where } D_{ij} > r_w \), \text{self-connections and existing mesh edges} 
        \ForAll{contact edges \( (i, j) \)}
            \State \( x_{ij}^c \leftarrow x_i^w - x_j^w \) \Comment{Relative position in contact space (except for the nodes from lines 10))}
            \State \( e_{ij}^c \leftarrow \text{CONCAT}(x_{ij}^c, \ |x_{ij}^c|) \) \Comment{Form contact edge features}
            \State \( e_{ij}^c \leftarrow \text{NORMALIZE}(e_{ij}^c) \) \Comment{Normalize features}
        \EndFor
    
    \State \Return \( v_i, e_{ij}^m, e_{ij}^c \)
    \end{algorithmic}
\end{algorithm}

In Algorithm~\ref{alg:mesh_to_graph}, we convert each element of the mesh into components suitable for graph representation: The mesh \( M \) comprises quadrilateral cell indices \( C_{\text{quad}} \), mesh coordinates \( x^m \), world coordinates \( x^w \), and node types \( n_i \). The mesh coordinates \( x^m \) represent the intrinsic positions within the mesh, preserving the internal structure over time. In contrast, the world coordinates \( x^w \) define the actual positions of nodes in the simulation space, which is crucial for modeling dynamics and interactions with the external environment. The node types \( n_i \) specify the function of each node \( i \), which can represent entities such as ball, display panel components, or boundary conditions. A contact radius \( r_w \) is also defined to establish contact edges between nearby nodes that are not directly connected in the mesh but can influence each other.

The converted graph structure is as follows: Node features  \( v_i \) are constructed by concatenating the one-hot encoded node type \( \varphi(n_i) \), the world space position \( x_i^w \), and the world space velocity \( \dot{x}_i^w \). Mesh edge features \( e_{ij}^m \) represent the connectivity and deformation of the mesh. These include the relative position in mesh space \( u_{ij} = x_i^m - x_j^m \) and its magnitude \( |u_{ij}| \), along with the relative position in world space \( x_{ij}^w = x_i^w - x_j^w \) and its corresponding magnitude \( |x_{ij}^w| \). Contact edge features \( e_{ij}^c \) used for interactions between nodes within the contact radius but not directly connected, include the relative position \( x_{ij}^c = x_i^w - x_j^w \) and its magnitude \( |x_{ij}^c| \).

A summary of these components is provided in Table~\ref{table:graph_structure}.

\begin{table}[H]
\centering
\renewcommand{\arraystretch}{1.3}
\begin{tabular}{ccc}
\hline
\textbf{Node Features} \( v_i \) & \textbf{Mesh Edge Features} \( e_{ij}^m \) & \textbf{Contact Edge Features} \( e_{ij}^c \) \\
\hline
\( \varphi(n_i), \ x_i^w, \ \dot{x}_i^w \) & \( u_{ij}, \ |u_{ij}|, \ x_{ij}^w, \ |x_{ij}^w| \) & \( x_{ij}^c, \ |x_{ij}^c| \) \\
\hline
\end{tabular}
\caption{Components of the Graph Structure \( G(V, E^M, E^C) \). The relative positions \( u_{ij} \) and \( x_{ij}^w \) are defined in mesh and world space, respectively, while \( x_{ij}^c \) represents contact interactions.}
\label{table:graph_structure}
\end{table}

By converting the mesh into this graph structure, we enable the MGN to learn complex dynamics by incorporating both the local mesh topology and potential interactions between nodes due to proximity. The mesh edges \( E^M \) represent the fixed connectivity of the mesh, preserving its structural integrity, while the contact edges \( E^C \) allow the model to account for interactions such as collisions or contact forces with other objects in the simulation. To fully utilize the rich information encoded in this graph structure, we employ an EPD architecture, which will be detailed in \Cref{sec:2.2}.

\vspace{10pt}
\subsection{Encoder-processor-decoder architecture}
\label{sec:2.2}
\vspace{10pt}

\begin{figure}[H]
    \centering
    \includegraphics[width=1\linewidth]{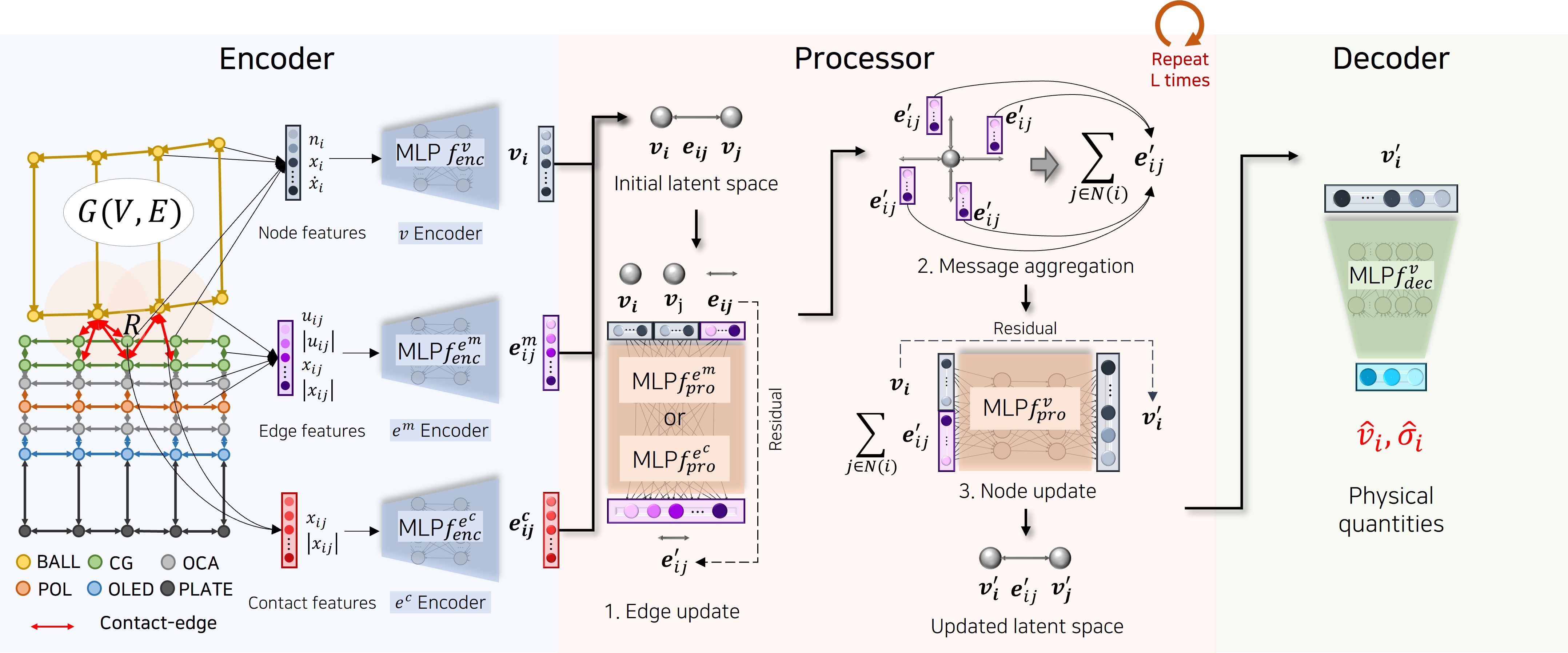}
    \captionsetup{justification=raggedright,singlelinecheck=false}
    \caption{Specialized MGN architecture for multi-layered display panel impact simulation.}
    \label{fig:Fig.2}
\end{figure}

As the foundation of our approach, we leverage the encoder-processor-decoder (EPD) architecture, as proposed by \cite{sanchez2020learning}, to model drop impact simulations, as illustrated in \cref{fig:Fig.2}. A comprehensive list of the materials used for the ball and multi-layer display panels can be found in \Cref{table:material_properties}. This architecture comprises three primary components: the encoder, the processor, and the decoder.

\textbf{Encoder structure} The encoder serves as the initial stage of our model, transforming the input graph structure $G(\textit{V}, E^M, E^C)$ into a high-dimensional latent space. This transformation is crucial for capturing the system's topological features and initial conditions. To perform this mapping, we utilize a multilayer perceptron (MLP), which enables us to learn rich representations of both nodes and edges. Specifically, the node features \(\textit{v}_i\) are derived through the application of this neural network, followed by a non-linear activation function and layer normalization (LN) \citep{ba2016layer}. The encoded node representation $\textit{v}_i$ is given by:

\begin{equation}
v_i = f_{enc}^v(v_{i}) = \text{LN}\left(\sigma\left(W^{(k)}v_i^{(k-1)} + b^{(k)}\right)\right)
\end{equation}
Similarly, the edge features, denoted as $e_{ij}^m$ and $e_{ij}^c$ for different types of edges, are encoded using the same mechanism. The encoded edge representations are expressed as:

\begin{equation}
e_{ij}^m = f_{enc}^{e^m}(e_{ij}^m) = \text{LN}\left(\sigma\left(W^{(k)}e_{ij}^{m^{(k-1)}} + b^{(k)}\right)\right)
\end{equation}

\begin{equation}
e_{ij}^c = f_{enc}^{e^c}(e_{ij}^c) = \text{LN}\left(\sigma\left(W^{(k)}e_{ij}^{c^{(k-1)}} + b^{(k)}\right)\right)
\end{equation}
where $f_{enc}^{e^m}$ and $f_{enc}^{e^c}$ are encoding functions implemented as networks that take $e_{ij}^m$ and $e_{ij}^c$, respectively. Here, $W^{(k)}$ and $b^{(k)}$ denote the weight matrix and bias vector of the $k$-th layer, respectively. $\sigma$ represents the ReLU activation function. LN refers to layer normalization, which is essential in this context as it stabilizes and accelerates the training process by normalizing the inputs across the features for each data point.

\begin{equation}
\text{LN}(x) = \frac{x-\text{E}[x]}{\sqrt{\text{Var}[x]+\epsilon}} \gamma + \beta
\end{equation}
where $\text{E}[x]$ and $\text{Var}[x]$ represent the mean and variance of the input features, respectively. $\epsilon$ is a small constant added for numerical stability, and $\gamma$ and $\beta$ are learnable parameters that allow the transformation to be rescaled and shifted. By incorporating LN and non-linear activation functions into the MLPs used in our architecture as well as the encoder, we can effectively capture the complex relationships within the graph structure, leading to more robust and generalizable representations of the data.

\textbf{Processor structure} The processor models the complex interactions and dynamics between nodes in the graph. It utilizes a message-passing scheme in which each node receives and processes messages (i.e., latent vectors) from its adjacent nodes connected by various edges (e.g., mesh edges, contact edges). This process is implemented using GNNs, which update the latent representations of nodes based on the information received.

As illustrated in \cref{fig:Fig.2}, the message-passing scheme within the processor can be divided into three primary steps:

\begin{enumerate}
    \item \textit{Edge update}. The edge update is represented by Eq. \ref{eq:5}. For each edge $(i,j)$, the updated edge $e'_{ij}$ is computed as:
    
    \begin{equation}
    e'_{ij} = f_{pro}^e(v_i, v_j, e_{ij}) = \text{LN}\left(\sigma\left(W^{(l)}\begin{bmatrix}v_i \\ v_j \\ e_{ij}\end{bmatrix}^{(l-1)} + b^{(l)}\right)\right)
    \label{eq:5}
    \end{equation}
    
    Here, $f_{pro}^e$ is the edge update function, which takes as input the features of nodes $v_i$ and $v_j$, and the current edge feature $e_{ij}$. As a results, the updated edges can allow the model to characterize the relationship between connected nodes.
    
    \item \textit{Message aggregation}. In Eq. \ref{eq:6}, for each node $v_i$, the updated edges connected to its neighboring nodes $v_j$ are aggregated:
    
    \begin{equation}
    \sum_{j\in N(i)} e'_{ij}
    \label{eq:6}
    \end{equation}
    
    This sum operation collects all the messages passed to node $i$ from its neighbors, preparing the information for the node update step.
    
    \item \textit{Node update}. Finally, the node update is shown as Eq. \ref{eq:7}:
    
    \begin{equation}
    v'_i = f_{pro}^v\left(v_i, \sum_{j\in N(i)} e'_{ij}\right) = \text{LN}\left(\sigma\left(W^{(l)}\begin{bmatrix}v_i \\ \sum_{j\in N(i)} e'_{ij}\end{bmatrix}^{(l-1)} + b^{(l)}\right)\right)
    \label{eq:7}
    \end{equation}
    
    Here, $f_{pro}^v$ is the node update function. It takes as input the current node feature $v_i$ and the aggregated messages from its adjacent nodes.
\end{enumerate}

The processor can consist of multiple message-passing layers (denoted by $l$ in the equations), allowing higher-order interactions and long-range dependencies between the nodes.

\textbf{Decoder structure} The decoder takes the latent representations of nodes, updated by the processor, and maps them back to the output space. Using a series of MLPs that act in the opposite way to the encoder, it predicts desired output quantities such as velocity $\hat{v}_i$ and stress $\hat{\sigma}_i$ for the subsequent time step. The decoder function can be expressed as:

\begin{equation}
\hat{v}_i, \hat{\sigma}_i = f_{dec}^v(v'_i) = \text{LN}\left(\sigma\left(W^{(l)}v_i^{(l-1)} + b^{(l)}\right)\right), \text{ except for LN of last layer } l
\end{equation}
The final layer omits LN to allow for unbounded output. 

\subsection{Physical constraint based loss function}
\label{sec:2.3}

\begin{figure}[H]
    \centering
    \includegraphics[width=1\linewidth]{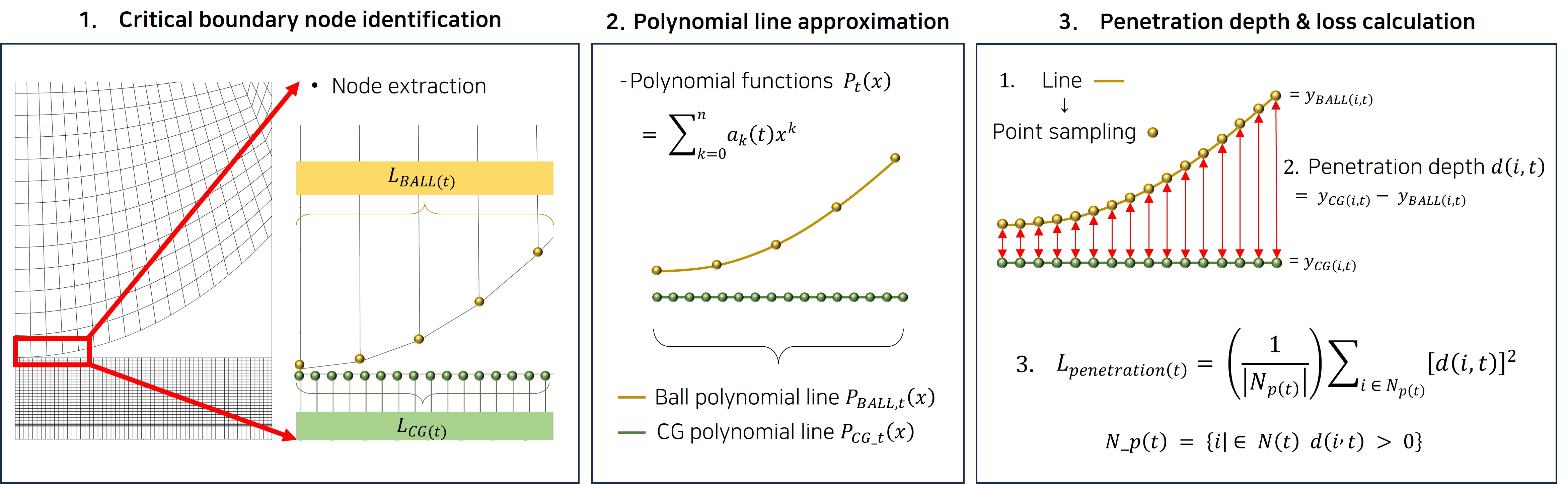}
    \captionsetup{justification=raggedright,singlelinecheck=false}
    \caption{Flowchart illustrating the incorporation of physical constraints in our model to mitigate inter-object penetration.}
    \label{fig:Fig.3}
\end{figure}

In this section, we introduce a physics-constrained loss function designed to improve the accuracy of predictive performance in drop impact scenarios (\cref{fig:Fig.3}). This approach specifically addresses the critical issue of unrealistic inter-object penetration, which not only violates fundamental principles of solid mechanics but also leads to inaccurate predictions. As a result, our approach can overcome the limitations of the data-driven spatio-temporal surrogate model proposed by \cite{Pfaff2020}, specifically specializing it for the drop impact scenarios and ensuring simulation validity.

To describe our physics-constrained loss function, we present a five-step process:

\mbox{}\\
\textit{\textbf{Step 1. Critical boundary node identification}} For each step in our trajectory data, which consists of 98 time steps, we identify two critical boundary lines:

\begin{itemize}
    \item Ball bottom boundary: $L_{BALL(t)} = \left\{(x, y) \in \mathbb{R}^2: y = \min_{\substack{i \in N_{BALL(t)} \\ x_i = x}} y_i\right\}$
    
    \item CG top boundary: $L_{CG(t)} = \left\{(x, y) \in \mathbb{R}^2: y = \max_{\substack{i \in N_{CG(t)} \\ x_i = x}} y_i\right\}$
\end{itemize}
where $N_{BALL(t)}$ and $N_{CG(t)}$ are the sets of nodes representing the ball and CG at time step $t$, respectively. In this formulation, for each $x$ coordinate, we extract the corresponding $y$ coordinate that represents the minimum (for the ball) or maximum (for the CG) value across all nodes with that specific $x$ coordinate. This ensures that we accurately identify the critical boundary nodes for each boundary line.

\mbox{}\\
\textit{\textbf{Step 2. Polynomial line approximation}} To efficiently represent the complex shapes of $L_{BALL(t)}$ and $L_{CG(t)}$ and accurately calculate their penetration depth, we approximate each boundary at time step $t$ using $n$th-degree polynomial functions $P_t(x)$:

\begin{equation}
P_t(x) = \sum_{k=0}^n a_k(t)x^k
\end{equation}
The choice of polynomial degree is critical, as it directly influences how well the boundary shapes are captured. Lower-degree polynomials may result in underfitting, failing to capture the complexity of the boundaries, while higher-degree polynomials might be too flexible, leading to overfitting. To achieve a high accuracy, we must carefully determine the optimal degree through parameter exploration. This parametric study, which examines the impact of different polynomial degrees on simulation accuracy, will be presented in \Cref{sec:Sec4}.

The coefficients $a(t) = (a_0(t), a_1(t), ..., a_n(t))^T$ are determined using least squares regression:

\begin{equation}
a(t) = (\textit{\textbf{V}}^T\textit{\textbf{V}})^{-1}\textit{\textbf{V}}^Ty(t)
\end{equation}

\begin{equation}
V = \begin{pmatrix}
1 & x_1 & \cdots & x_1^n \\
1 & x_2 & \cdots & x_2^n \\
\vdots & \vdots & \ddots & \vdots \\
1 & x_m & \cdots & x_m^n
\end{pmatrix}
\end{equation}
where \textit{\textbf{V}} is the Vandermonde matrix and $y(t)$ are the observed $y$-coordinates at time step $t$. The Vandermonde matrix \textit{\textbf{V}} is structured with each row representing powers of $x$ from 0 to $n$, facilitating polynomial fitting through least squares regression. This regression method is chosen for its ability to minimize the sum of squared residuals, providing a smooth approximation that captures the overall shape of the boundaries while reducing the impact of potential noise in the node positions. With these polynomial approximations $P_t(x)$ for both $L_{BALL(t)}$ and $L_{CG(t)}$, we can now accurately calculate the penetration depth between the ball and the CG at each time step, which is crucial for our penetration loss calculation.

\mbox{}\\
\textit{\textbf{Step 3. Penetration depth calculation}} The penetration depth is calculated by the difference between the $y$-values of the ball and CG boundaries. Let $y_{BALL(i,t)}$ and  $y_{CG(i,t)}$ denote the $y$-values of the polynomial functions $P_t(x)$ at the $i$-th node for the ball and CG lines at time step $t$, respectively. The penetration depth $d(i,t)$ at a given $i$-th node and time step $t$ is given by:

\begin{equation}
d(i,t) = y_{CG(i,t)} - y_{BALL(i,t)}
\end{equation}
If $d(i,t)$ has positive values, it indicates the inter-object penetration, where the ball's surface has penetrated the CG's surface at that specific time step. Negative values indicate there is no contact between the ball and the CG.

\mbox{}\\
\textit{\textbf{Step 4. Penetration loss function}} We introduce a penetration loss term $L_{penetration}$ based on the mean squared error (MSE) of positive penetration depths. For the time step $t$, its penetration loss is calculated as:

\begin{equation}
L_{penetration(t)} = \left(\frac{1}{|N_{p(t)}|}\right) \sum_{i \in N_{p(t)}} [d(i,t)]^2
\end{equation}
where $N_{p(t)}$ is the set of nodes with positive penetration at time step $t$, i.e., $N_p(t) = \{i \in N(t) | d(i,t) > 0\}$, and $|N_{p(t)}|$ is the number of such nodes. If there are no positive penetrations at time step $t$, we set $L_{penetration(t)} = 0$. The minimization of such a loss term is expected to hinder the degree of penetration since $d(i,t)$ is expected to decrease.

\mbox{}\\
\textit{\textbf{Step 5. Integration into total loss function}} The final loss function $L_{total(t)}$ is a weighted sum of the data-driven loss $L_{data(t)}$ (i.e., the sum of velocity and stress loss) and the penetration loss in a single time step $t$:

\begin{equation}
L_{total(t)} = L_{data(t)} + \lambda \cdot L_{penetration(t)}
\end{equation}
where $\lambda > 0$ is a hyperparameter controlling the contribution of the penetration loss. In \Cref{sec:Sec4}, we also present the results of a parametric study on $\lambda$, examining several case studies to determine the optimal value that balances penetration prevention and overall simulation accuracy.

\section{Data generation and model training}
\label{sec:implemetation details}

\subsection{Data generation}
\label{subsec:sec3.1}

\begin{figure}[H]
    \centering
    \includegraphics[width=1\linewidth]{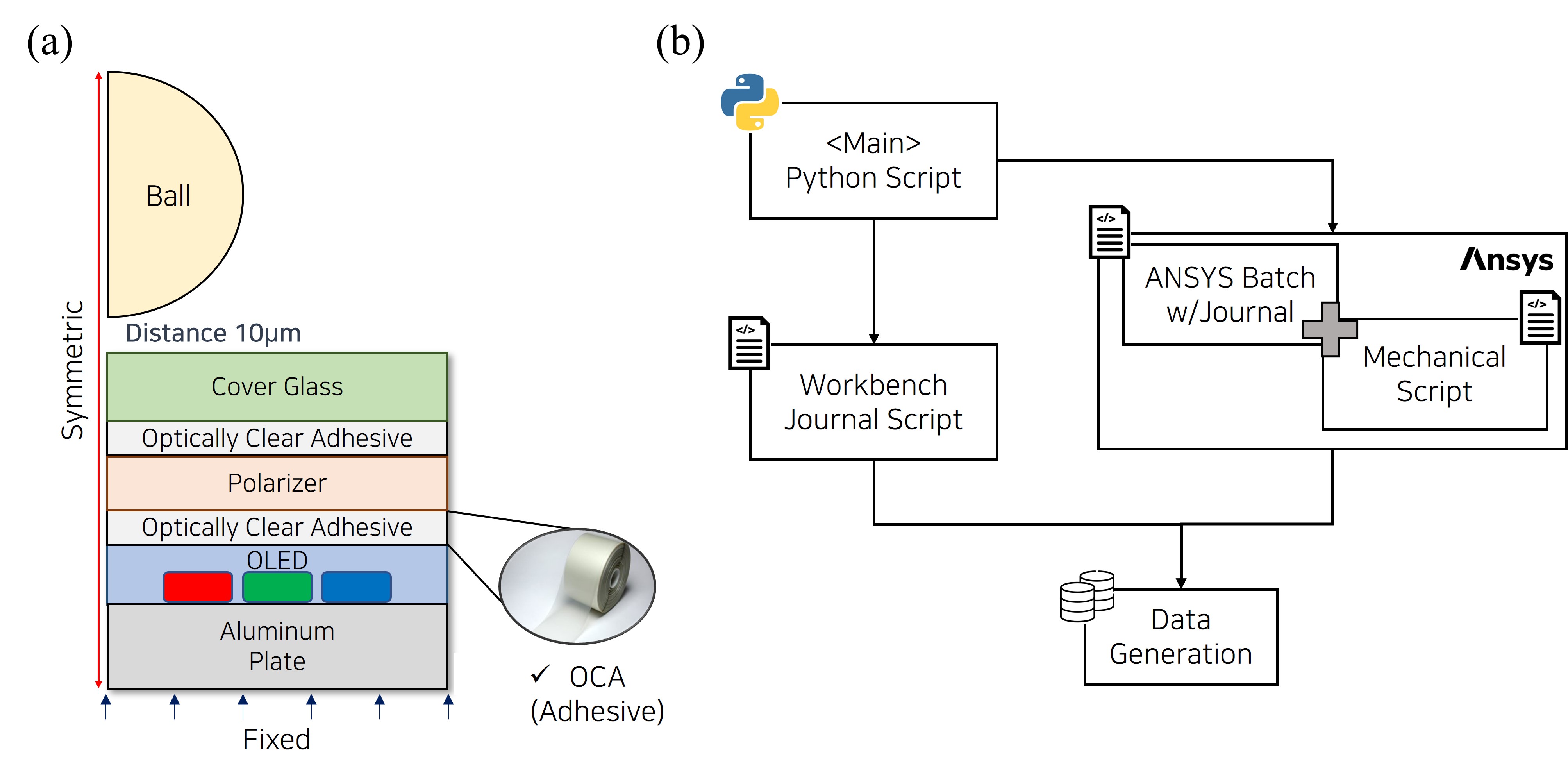}
    \captionsetup{justification=raggedright,singlelinecheck=false}
    \caption{Schematic of the data generation process: (a) Material structure used in impact test, (b) Automated data generation process.}
    \label{fig:Fig.4}
\end{figure}

The dataset used in this study is structured as shown in \cref{fig:Fig.4}(a) and the properties of each material are listed in \Cref{table:material_properties} \citep{Corning2020, Ma2020, Wu2022}. In this study, the material model is assumed that the OCA behaves as a linear material, rather than using a hyperelastic material model. Based on this assumption, the dataset is generated to include different mesh configurations with different thickness of the OCA, while the initial location of the ball above the plate is fixed at 10\textmu m. The reason for this height is to alleviate computational burden by removing the unnecessary time steps to be solved by the computer simulation --- therefore the impact velocity is maintained at 0.62631 m/s for all datasets. The overall process of dataset generation is illustrated in \cref{fig:Fig.4}(b). Specifically,the data used for training and evaluation is obtained using the explicit dynamics solver in ANSYS Workbench software \citep{ansys_workbench_2024}.  

\begin{table}[H]
\centering
\tabularfont
\begin{adjustbox}{max width=\textwidth}
\begin{tabular}{cccc}
\hline
Layer & \begin{tabular}[c]{@{}c@{}}Elastic\\Modulus \lbrack GPa\rbrack\end{tabular} & \begin{tabular}[c]{@{}c@{}}Poisson's\\Ratio \end{tabular} & \begin{tabular}[c]{@{}c@{}}Thickness\\\lbrack$\mu$m\rbrack\end{tabular} \\
\hline
Ball & 200 & 0.3 & 5,000 (radius) \\
Cover glass (CG) & 77 & 0.21 & 100 \\
Optically clear adhesive 1 ($\text{OCA}_1$) & 0.01 & 0.45 & 50$\sim$150 \\
Polarizer (POL) & 4 & 0.33 & 50 \\
Optically clear adhesive 2 ($\text{OCA}_2$) & 0.01 & 0.45 & 50$\sim$150 \\
organic light emitting diodes (OLED) & 5.15 & 0.3 & 30 \\
Aluminum plate (PLATE) & 68.9 & 0.33 & 1,200 \\
\hline
\end{tabular}
\end{adjustbox}
\captionsetup{justification=raggedright, singlelinecheck=false}
\caption{Material properties of the dataset.}
\label{table:material_properties}
\end{table}

To efficiently use mesh elements for training with limited GPU memory, we simplified the mesh data from 11,526$\sim$14,438 nodes to 2,403$\sim$3,117 nodes, focusing on the region where fluctuations in value occur during the ball's impact on the plate. As shown in \cref{fig:Fig.4}, the boundary conditions include a fixed constraint applied to the bottom of the PLATE. Additionally, a symmetric condition was imposed to significantly reduce analysis time --- by exploiting the symmetry of the model, only right part of the structure needed to be analyzed.

\begin{figure}[H]
    \centering
    \includegraphics[width=1\linewidth]{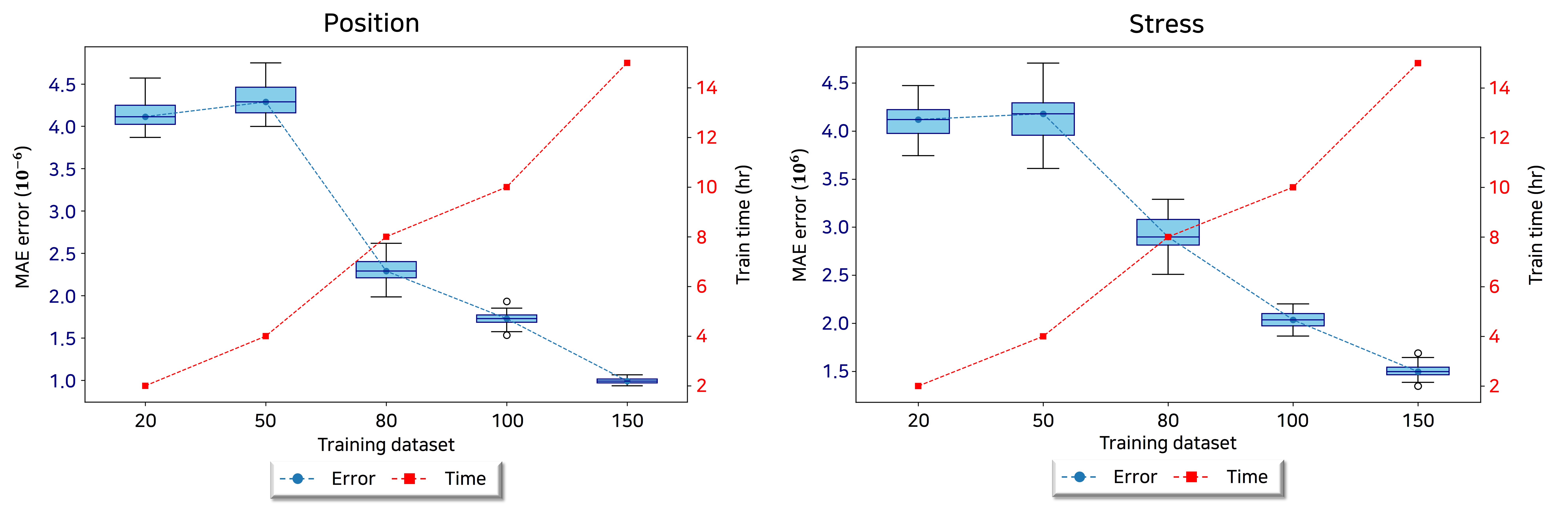}
    \captionsetup{justification=raggedright,singlelinecheck=false}
    \caption{Validation of model accuracy and efficiency based on the number of training datasets.}
    \label{fig:Fig.5}
\end{figure}

Under these explicit dynamics conditions, a time interval of 4$\mu$s was used in the experiments for a total of 100 time steps. To ensure comprehensive design space, we selected design points using Latin hypercube sampling when designing the experiment \citep{mckay2000comparison}. Then, to consider the balance between accuracy and efficiency in our model when splitting train-test trajectories, we calculated the average performance across five MGN models with varying data size, 20, 50, 80, 100, 150 trajectories for the training data, as shown in \cref{fig:Fig.5}. From these results, it is observed the error decreases significantly as the data size increases until 80 trajectories. However, the model still shows higher error fluctuations. Therefore, by considering both the decrease in error and the increase in training time, we determined that 100 trajectories offer an good balance between model accuracy and efficiency, making it the appropriate choice for the training dataset.

\subsection{Model settings}
\subsubsection{Model hyperparameters}
The networks used in the encoder, processor, and decoder are implemented with MLPs. Each MLP consists of two hidden layers with ReLU activation functions and maintains a 128-dimensional latent vector. The model is trained using the Adam optimizer with a learning rate of 0.001 and a batch size of 1, for 100 epochs. The number of message-passing layers in the processor is set to 15, which is judged to be sufficient for capturing higher-order interactions between the nodes.

\subsubsection{Model inference}
During model inference, we can predict one future time step (in a process called `rollout') with respect to a full-step simulation by utilizing the initial state mesh data from unseen trajectories. In a step-by-step rollout prediction, the forward Euler method is applied to obtain the position of the objects at the subsequent time step, as shown in Eq. \ref{eq:model inference}:

\begin{equation}
x_i(t + 1) = x_i(t) + \hat{\dot{x}}_i\Delta t
\label{eq:model inference}
\end{equation}
Here, $x_i(t + 1)$ represents the position of node $i$ at time step $t + 1$, $x_i(t)$ denotes the position of node $i$ at time step $t$, $\hat{\dot{x}}_i$ indicates the predicted velocity of node $i$ by the MGN models, and $\Delta t$ is the time step size, which is fixed as 4$\mu$s in \Cref{subsec:sec3.1}. In \Cref{sec:Sec4}, we will further explore the impact of using methods other than forward Euler for rollout prediction.

\subsubsection{Model implementation and message-passing scheme}
Our model is implemented using the PyTorch deep learning framework \citep{paszke2019pytorch}, and the experiments are conducted on an NVIDIA GeForce RTX 3090 GPU with 24GB of memory. The message-passing layers were implemented using a gather-scatter scheme, employed in PyTorch Geometric \citep{fey2019fast}. This scheme exhibits non-deterministic behavior due to the use of GPU atomic operations, which can lead to varying results across multiple training and testing runs, even when the same random seed is used. While a fully-deterministic implementation is possible by directly utilizing sparse matrix multiplications, it appears to be significantly slower compared to the non-deterministic approach \citep{gao2024finite}. Therefore, our study also adopted the non-deterministic characteristic to prioritize computational efficiency.

\section{Prediction results}
\label{sec:Sec4}

\subsection{Limitations of vanilla MGN}

\begin{figure}[H]
    \centering
    \includegraphics[width=0.9\linewidth]{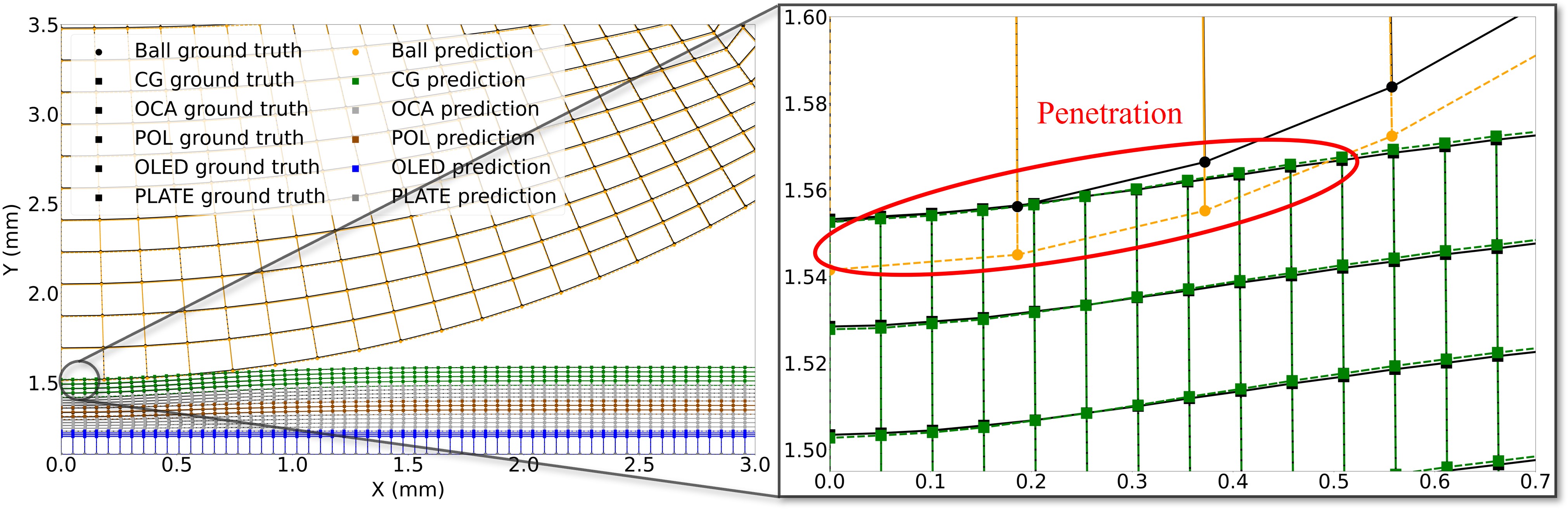}
    \captionsetup{justification=raggedright,singlelinecheck=false}
    \caption{The prediction performance of a vanilla MGN for ball drop impact on a multi-layered display panel. (Left) Full prediction view. (Right) Zoomed-in view highlighting penetration (red circle) between ball and CG layer. Solid lines/black marker represent ground truth; dashed lines/color markers show predictions.}
    \label{fig:Fig.6}
\end{figure}

MeshGraphNets (MGNs), proposed by \citet{Pfaff2020}, offer a versatile graph neural network model applicable to various physical systems. Vanilla MGNs excel in handling unstructured mesh data flexibly and allow for the consideration of interactions between objects through edge connections based on spatial proximity. In this section, we show the performance of vanilla MGN in predicting the nonlinear behavior caused by the drop impact of a ball falling onto multi-layered display panels (\cref{fig:Fig.6}). Specifically, we aim to predict the deformations and stresses that occur over time in both flexible bodies. 

As \cref{fig:Fig.6}, a vanilla MGN fails to maintain physical constraints, resulting in penetration between the ball and the CG layer. It highlights the need for improved approaches for vanilla MGN model that preserve physical consistency while leveraging the computational advantages of GNNs. To address these limitations and improve the accuracy of predictions, in the following section we propose physics-constrained MGN approach specialized for ball drop impact test, and additionally conduct the parametric study to optimize its architecture, ultimately aiming to overcome the drawbacks observed in the vanilla MGN.

\vspace{10pt}
The key problems of penetration encountered by vanilla MGN are:

\begin{itemize}
    \item \textit{Violation of physical laws}: Impact simulations should satisfy the momentum/energy conservation laws. However, penetration indicates that the surrogate model or contact algorithm fails to enforce these principles correctly. This failure can lead to predictions of post-impact behavior that significantly deviates from physical reality.
    \item \textit{Inaccurate stress distribution}: Penetration results in the prediction model inaccurately estimating stress distributions due to incorrect representation of contact interactions. This inaccuracy leads to erroneous predictions of object behavior and failure modes, undermining the reliability of the model's outcomes.
    \item \textit{Cumulative error propagation}: Penetration errors at the initial time steps can propagate through subsequent time steps. This error accumulation leads to progressively divergent predictions, severely undermining the model's long-term predictive capabilities.
\end{itemize}

\subsection{Parametric study to improve the accuracy of physics-constrained model}
\subsubsection{Exploring polynomial degree used in physics-constrained approach}
In \Cref{sec:2.3}, we proposed an approach to prevent penetration phenomena between flexible bodies. To calculate penetration depth accurately, we applied the line approximation technique, approximating both the ball and plate boundaries with polynomial functions. In this section, we further explore how the complexity of boundary shapes influences model performance in this section. Boundary representations are denoted as B.n$_1$/P.n$_2$, where n$_1$ and n$_2$ represent the polynomial degrees used to approximate the boundaries of the ball and display panel geometries, respectively. For example, B.2/P.3 denotes that 2nd order and 3rd order polynomials are used for modeling the shape of ball and display panel, respectively. And the selection of polynomial degrees can be guided by geometric characteristics and expected deformation patterns during impact. Even-degree polynomials (2, 4, and 6) are used for the ball to represent its initial spherical shape and potential deformations during impact. Odd-degree polynomials (3, 5, and 7) are chosen for the panel to account for its initially flat state and to capture the asymmetric bending modes that typically occur during impact. These combinations of ball and display panel approximations are determined to result in a total of 9 cases, which are conducted to assess their impact on model performance: results of these cases are summarized in \Cref{table:poly_degree}.

\begin{table}[H]
\centering
\tabularfont
\renewcommand{\arraystretch}{1.3} 
\begin{adjustbox}{max width=\textwidth}
\begin{tabular}{c|cccccccc}
\hline
\multicolumn{1}{c|}{} & \multicolumn{8}{c}{\textbf{Polynomial degree}} \\
\cline{2-9}
& B.2/P.3 & B.2/P.5 & B.2/P.7 & B.4/P.3 & B.4/P.5 & B.4/P.7 & \textbf{B.6/P.5} & B.6/P.7 \\
\hline
Position MAE \lbrack$10^{-6}$\rbrack & 1.420 & 0.539 & 1.273 & 0.471 & 0.626 & 0.909 & \textbf{0.309} & 0.545 \\
Position RMSE \lbrack$10^{-6}$\rbrack & 5.486 & 1.131 & 4.917 & 1.036 & 1.953 & 2.931 & \textbf{0.707} & 1.288 \\
Stress MAE \lbrack$10^{6}$\rbrack & 1.314 & 1.111 & 1.276 & 1.033 & 1.162 & 1.054 & \textbf{0.547} & 0.718 \\
Stress RMSE \lbrack$10^{6}$\rbrack & 8.664 & 5.362 & 8.583 & 5.188 & 2.228 & 8.164 & \textbf{2.223} & 3.052 \\
Time [h] & 21.917 & 22.017 & 21.750 & \textbf{21.133} & 21.583 & 21.917 & 21.233 & 21.283 \\
\hline
\end{tabular}
\end{adjustbox}
\captionsetup{justification=raggedright, singlelinecheck=false}
\caption{Model performance across various polynomial degree combinations of flexible body boundaries in drop impact prediction is evaluated.}
\label{table:poly_degree}
\end{table}

Our results demonstrate that the decision of polynomial degrees significantly affects the model's capability to precisely predict drop impact behaviors. In this context, MAE refers to the mean absolute error, and RMSE stands for root mean squared error. The B.6/P.5 configuration exhibits superior performance, achieving the lowest errors among all configurations. Compared to the configuration with the second-lowest errors (B.4/P.5), the B.6/P.5 model shows improvements of 50.6\% in position MAE and 63.8\% in position RMSE. A key point in this section is that providing higher capacity to the polynomial by increasing its degree does not necessarily enhance performance. For instance, increasing the panel's polynomial degree from 5 to 7 (B.6/P.7) rather leads to a decline in accuracy with position MAE increasing by 76.4\% and stress MAE by 31.3\%. This suggests an optimal complexity threshold for our approximations, beyond which the model may suffer from overfitting or numerical instability. Computational time remains almost consistent across all configurations, with only a 4.2\% variation. This consistency in computational cost allows us to prioritize maximizing accuracy without trade-offs in efficiency when selecting polynomial degrees.

\subsubsection{The effect of penetration loss weight on physics constraints}

\begin{table}[H]
\centering
\tabularfont
\renewcommand{\arraystretch}{1.3} 
\begin{adjustbox}{max width=\textwidth}
\begin{tabular}{c|ccccc}
\multicolumn{6}{r}{\textsuperscript{1}Without penetration loss (Vanilla MGN)} \\
\hline
\multicolumn{1}{c|}{} & \multicolumn{5}{c}{\textbf{Penetration Loss Weight $\lambda$}} \\
\cline{2-6}
 & None$^1$ & 0.1 & 1 & 10 & 30 \\
\hline
Position MAE \lbrack$10^{-6}$\rbrack & 0.704 & 0.783 & 0.830 & \textbf{0.309} & 0.583 \\
Position RMSE \lbrack$10^{-6}$\rbrack & 2.509 & 2.548 & 2.384 & \textbf{0.707} & 1.566 \\
Stress MAE \lbrack$10^{6}$\rbrack & 0.9775 & 0.826 & 0.717 & \textbf{0.547} & 0.964 \\
Stress RMSE \lbrack$10^{6}$\rbrack & 11.744 & 4.188 & 2.773 & \textbf{2.223} & 4.917 \\
Time [h] & \textbf{19.617} & 21.617 & 21.283 & 21.233 & 21.9 \\
\hline
\end{tabular}
\end{adjustbox}
\captionsetup{justification=raggedright, singlelinecheck=false}
\caption{Validation of varying penetration loss weights in drop impact prediction.}
\label{table:lambda_effect}
\end{table}

We examine the influence of penetration loss weight ($\lambda$) on the model performance as summarized in \Cref{table:lambda_effect}. When $\lambda$ is set to 10, the model achieves optimal results across all metrics, yielding the lowest errors. Compared to the baseline model without penetration loss ($\lambda = \text{None}$), the optimal configuration demonstrates significant improvements: 56.1\% reduction in position MAE, 71.8\% reduction in position RMSE, 44.0\% reduction in stress MAE, and 81.1\% reduction in stress RMSE. 

However, increasing $\lambda$ beyond this optimal point does not necessarily enhance  performance. When $\lambda$ is increased to 30, there is an 88.7\% increase in position MAE and a 76.3\% increase in stress MAE compared to $\lambda = 10$. This suggests that excessively strict penetration constraints can disrupt the balance between the data-driven and physics-constrained loss function, leading to a degradation in overall prediction accuracy. This imbalance may result from the model overemphasizing the physical constraint (penetration) at the expense of capturing the inherent global dynamics of the two objects, which ultimately compromises the global accuracy of the impact test.

\begin{figure}[H]
    \centering
    \includegraphics[width=0.9\linewidth]{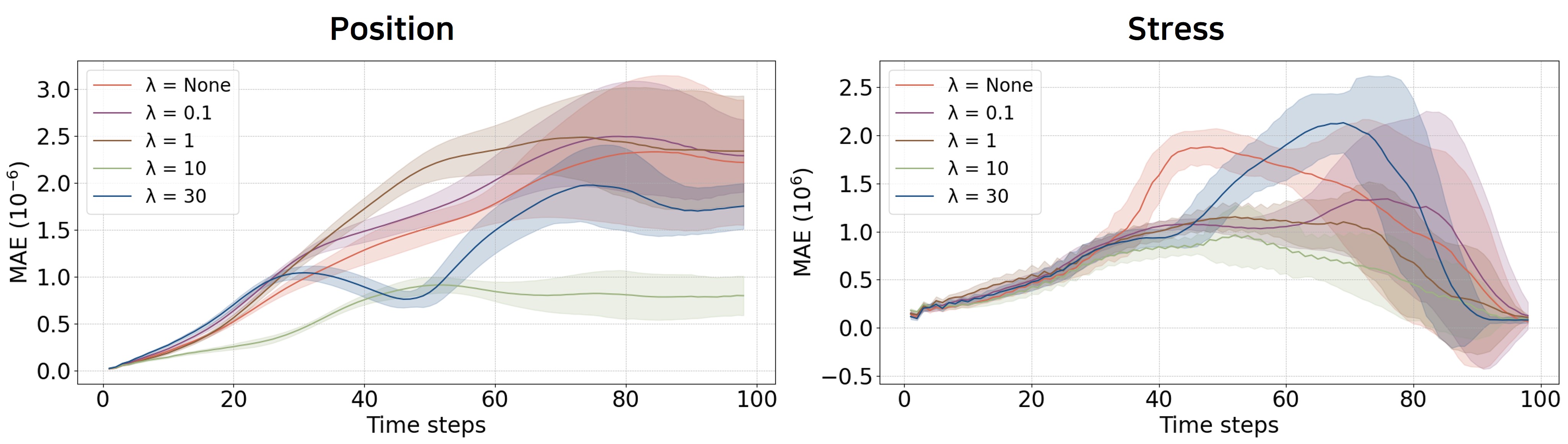}
    \captionsetup{justification=raggedright,singlelinecheck=false}
    \caption{For different penetration loss weights ($\lambda$), the mean and standard deviation of the MAE for position (left) and stress (right) predictions at each time-step rollout, based on all test trajectories.}
    \label{fig:Fig.7}
\end{figure}

For the detailed invesgigation on the rollout performance of the trained models, we show the mean and standard deviation of the MAE values at each time step for the 50 test trajectories in \cref{fig:Fig.7}. As mentioned in Eq. \ref{eq:model inference}, the model makes step-by-step predictions in the inference stage, which can lead to error accumulation over time. According to \cref{fig:Fig.7}, the configuration with $\lambda=10$ demonstrates the most stable performance for both position and stress predictions across the all time steps. Specifically, for position predictions, the configuration with $\lambda=10$ exhibits consistently lower overall error accumulation. Although stress prediction errors are comparable across all $\lambda$ values in the initial stages, $\lambda=10$ shows superior long-term performance. This stability is evidenced by the consistently lower MAE and narrower standard deviation band throughout the simulation. In contrast, The other cases show a rapid error accumulation with higher variability than $\lambda=10$ case. This suggests that $\lambda = 10$ is the optimal setting for balancing accuracy and stability in long-term predictions. 

\begin{figure}[H]
    \centering
    \includegraphics[width=0.9\linewidth]{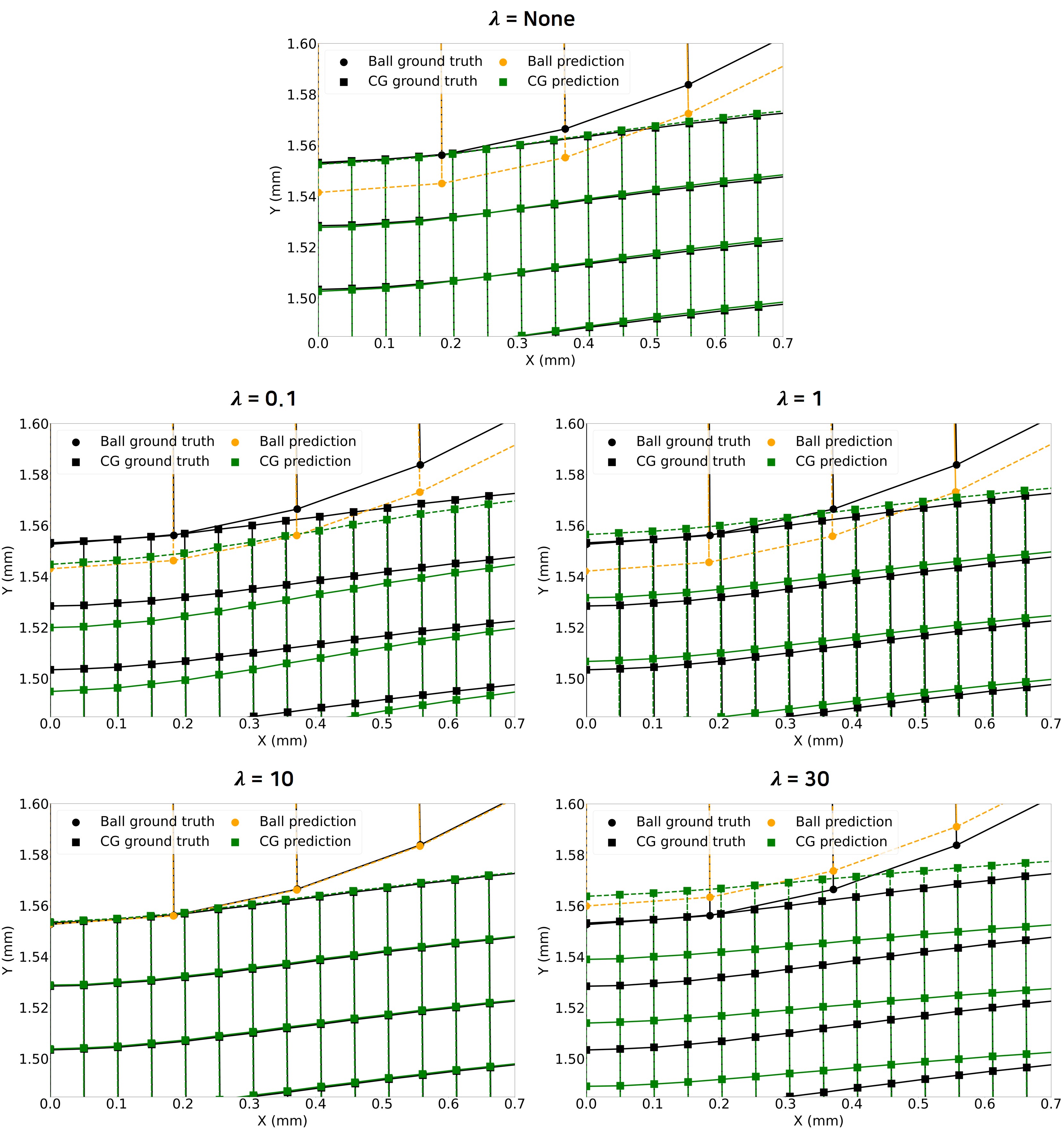}
    \captionsetup{justification=raggedright,singlelinecheck=false}
    \caption{Visualization of position predictions at the moment of ball rebound for different penetration loss weights ($\lambda$). Solid lines/black marker represent ground truth; dashed lines/color markers show predictions.}
    \label{fig:Fig.8}
\end{figure}

\cref{fig:Fig.8} illustrates how the model predicts positions at the moment of ball rebound across various $\lambda$ values. When $\lambda=0.1$, penetration between the ball and the CG layer is reduced compared to $\lambda=\text{None}$, but both objects rebound later than the ground truth due to insufficient penetration constraint on the system's dynamics. At $\lambda=1$, the model's predictions remain misaligned, with the ball rebounding too late and the panels slightly early, indicating an imbalance in the constraints. With $\lambda=30$, penetration is quite mitigated, but both objects are predicted to rebound faster than the ground truth, leading to a loss of accuracy. In contrast, with $\lambda=10$, no penetration occurs, and the accuracy closely aligns with the ground truth. This configuration achieves a desirable balance between minimizing local penetration and preserving global dynamics, making it the most suitable choice for this task. Therefore, we conclude that $\lambda=10$ offers the best compromise between preventing penetration and maintaining overall prediction accuracy.

\subsection{Comprehensive comparison between vanilla MGN and physics-constrained MGN in predicting impact behavior}

This section presents the performance comparison between the vanilla MGN and our physics-constrained MGN models in predicting the behavior of multi-layered display panels subjected to ball drop impact. To further clarify the dynamics of these impact systems, we first summarize the behavior of two flexible bodies across various time phases.

\begin{itemize}
    \item Phase 1: Ball drop, no contact
    \item Phase 2: Ball drop, contact
    \item Phase 3: Ball rebound, contact
    \item Phase 4: Ball rebound, no contact
\end{itemize}

\begin{figure}[htb!]
    \centering
    \includegraphics[width=0.9\linewidth]{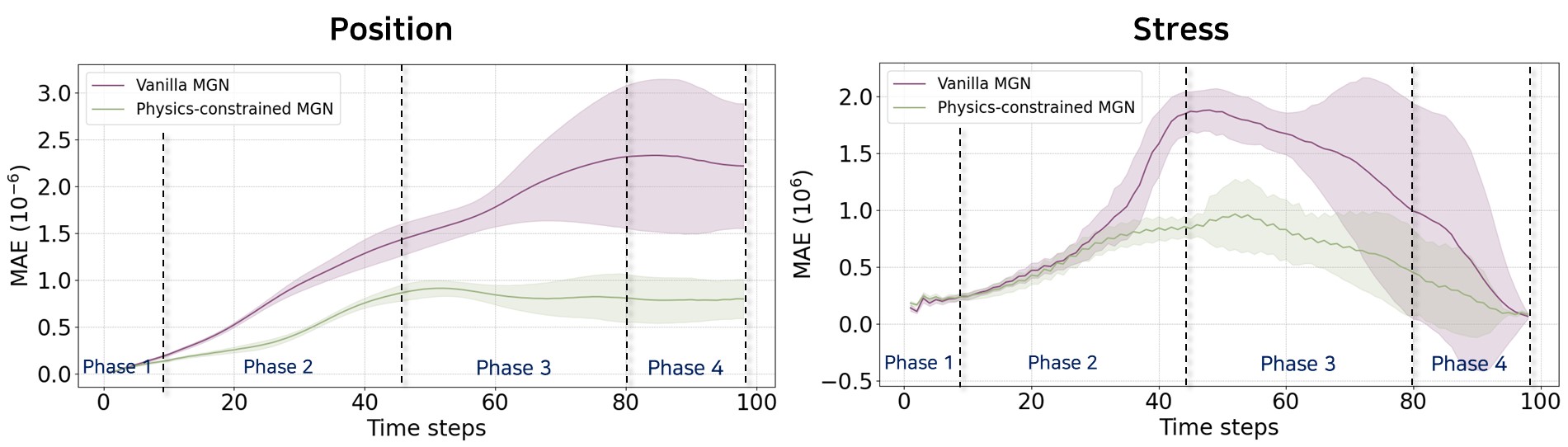}
    \captionsetup{justification=raggedright,singlelinecheck=false}
    \caption{Rollout-based performance predictions of vanilla MGN and physics-constrained MGN, averaged at each time-step across all test trajectories. This figure represents differences in MAE for both position (left) and stress (right) predictions across various time phases, including drop, impact, and rebound stages.}
    \label{fig:Fig.9}
\end{figure}

\cref{fig:Fig.9} demonstrates the error accumulation in accordance with step-by-step prediction. Overall, while both models exhibit increasing errors over time, the physics-constrained MGN demonstrates significantly lower error accumulation rates and reduced variability compared to the vanilla MGN. Specifically, the vanilla MGN struggles to accurately model the position of objects due to penetration by drop impact. This results in a noticeable divergence and rapidly increasing MAE, with continuous error accumulation, particularly during the rebound phase. Consequently, the vanilla MGN suffers from extreme stress predictions with a significantly higher peak and pronounced fluctuations when the maximum contact occurs (between phase 2 and phase 3) due to the inferior performance at accurately modeling the non-penetration dynamics, ultimately failing to restore back to its original position during the rebound phase.

In contrast, our proposed model shows superior predictive performance during drop impact, largely attributable to the penetration constraint. This is clearly reflected in the position performance, where the physics-constrained MGN shows much less error with respect to position, maintaining a consistently low MAE throughout the contact and rebound phases. Moreover, it presents a moderate stress peak at impact, with a slight increase in stress during contact and minimal error accumulation during the rebound phase compared to vanilla MGN. By examining this divergence, we can indirectly confirm that the physics-constrained model is significantly more effective at preventing penetration compared to its vanilla counterpart.

\begin{figure}[htb!]
    \centering
    \includegraphics[width=0.9\linewidth]{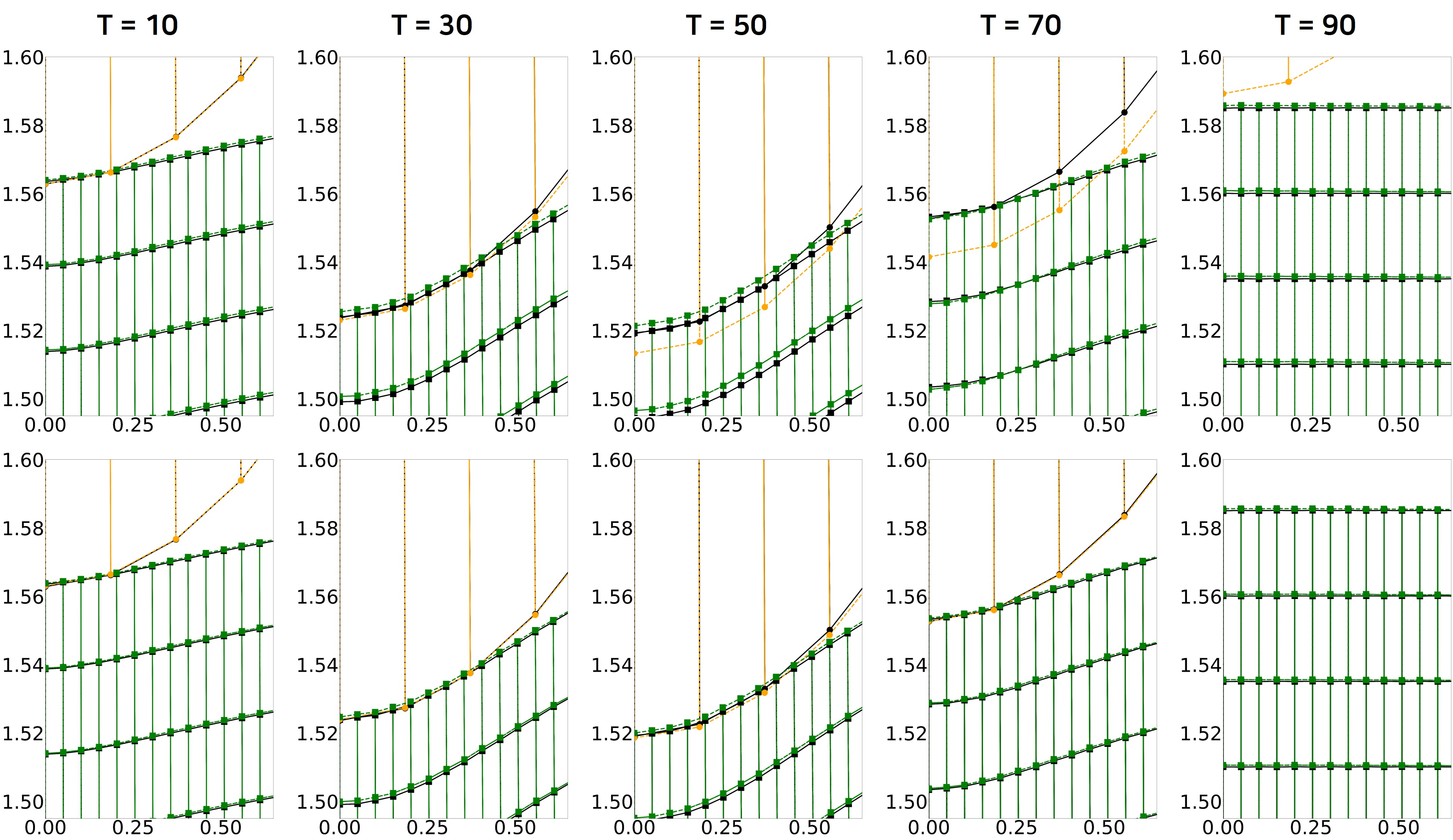}
    \captionsetup{justification=raggedright,singlelinecheck=false}
    \caption{Visualization for the position prediction at time steps (top: Vanilla MGN; bottom: Physics-constrained MGN; Axes units: $10^{-3}$).}
    \label{fig:Fig.10}
\end{figure}

\begin{figure}[htb!]
    \centering
    \includegraphics[width=0.9\linewidth]{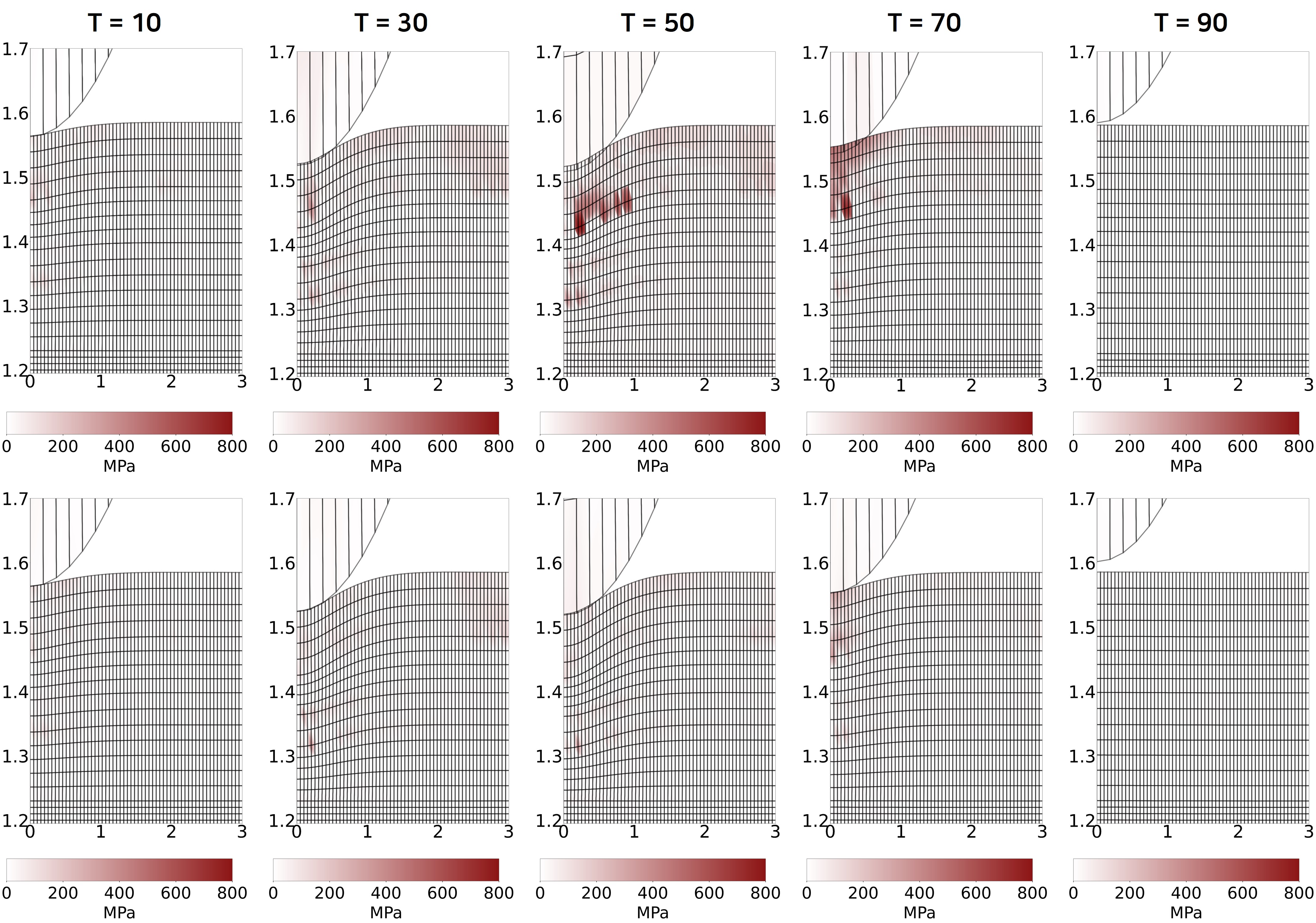}
    \captionsetup{justification=raggedright,singlelinecheck=false}
    \caption{Visualization of stress error predictions at each time step (top: Vanilla MGN; bottom: Physics-constrained MGN).}
    \label{fig:Fig.11}
\end{figure}

We intuitively compare the rollout prediction results of the vanilla MGN and our physics-constrained MGN based on position error (\cref{fig:Fig.10}) and stress error (\cref{fig:Fig.11}). Initially, at T=10, both models show comparable accuracy in position prediction. However, as time progresses, significant errors emerge, revealing the strengths of our proposed approach. Specifically, the vanilla MGN shows a gradual decline in accuracy, particularly after T=50, primarily due to penetration issues that result in delayed predictions. This growing discrepancy is clearly illustrated in the zoomed-in view at later time steps. In contrast, physics-constrained MGN consistently maintains a high level of accuracy throughout the entire sequence, with minimal drift from the ground truth. Furthermore, by examining the absolute difference between predicted and ground truth stress $|\sigma - \hat{\sigma}|$ (\cref{fig:Fig.11}), we observe that our proposed model achieves substantially lower stress errors across all time steps. In particular, it can be seen that the vanilla MGN model shows a clear trend of increased error at T=30, 50, where the penetration occurs --- while the proposed model shows reasonable errors at the same time steps but with relatively much lower stress error than the vanilla one.

\begin{figure}[htb!]
    \centering
    \includegraphics[width=0.9\linewidth]{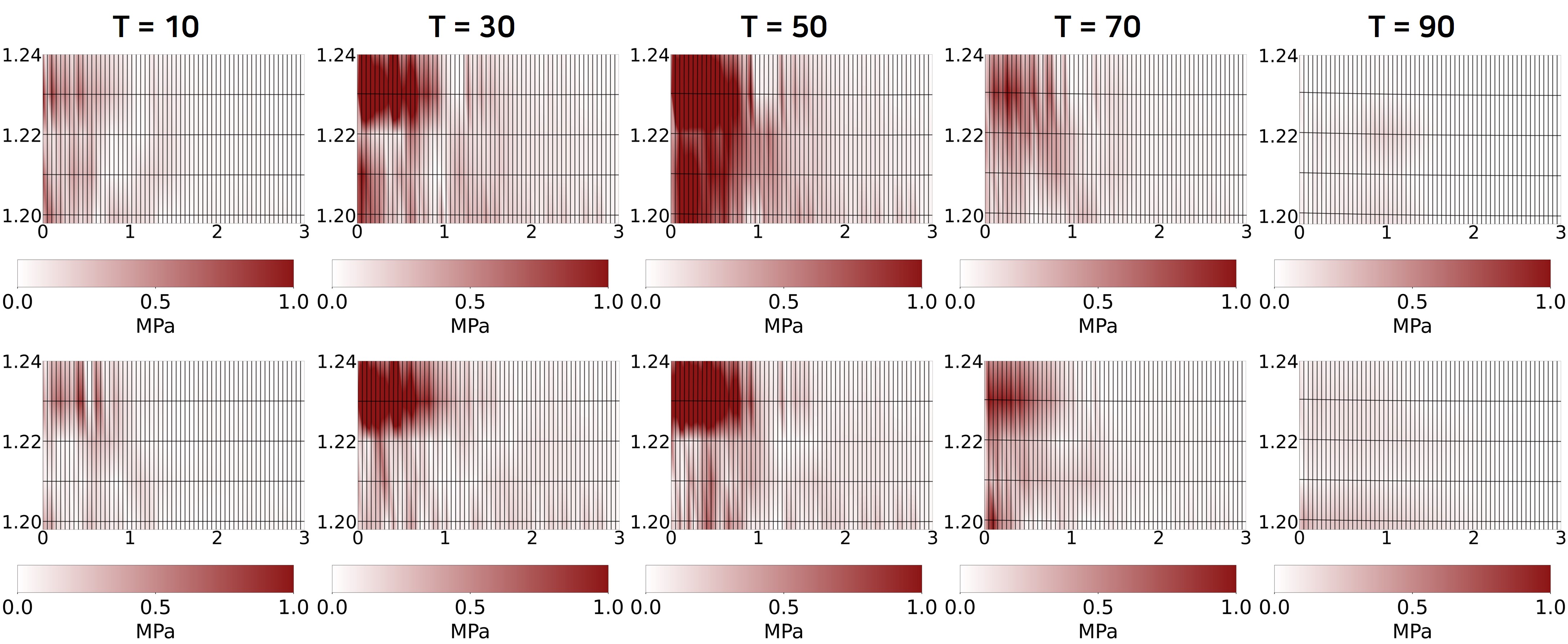}
    \captionsetup{justification=raggedright,singlelinecheck=false}
    \caption{Visualization of stress error predictions in OLED structure at each time step (top: Vanilla MGN; bottom: Physics-constrained MGN.}
    \label{fig:Fig.12}
\end{figure}

\cref{fig:Fig.12} focuses on a stress error distribution over time especially for OLED material, a critical light-emitting component in multi-layered display panels. Physics-constrained MGN still shows superior performance compared to the vanilla MGN model, exhibiting lower stress errors across time steps. This improved accuracy is crucial for optimizing display design, where the thickness of the OCA layer is optimized to minimize stress on the OLED material. In this regard, accurate stress prediction is essential for maintaining the longevity and performance of OLED displays, as even minor errors can lead to suboptimal designs and reduced durability. The improved precision of our model, reflected in \cref{fig:Fig.12} by the more localized and reduced distribution of the red areas in the lower panels, especially at T=50, 70 where the maximum stress is generated, enables more reliable prediction of the stress distribution prediction. In \Cref{sec:Sec5}, we will delve into the process of optimizing OCA thickness using our high-dimensional prediction model, which builds upon spatio-temporal modeling techniques.

\begin{figure}[htb!]
    \centering
    \includegraphics[width=0.9\linewidth]{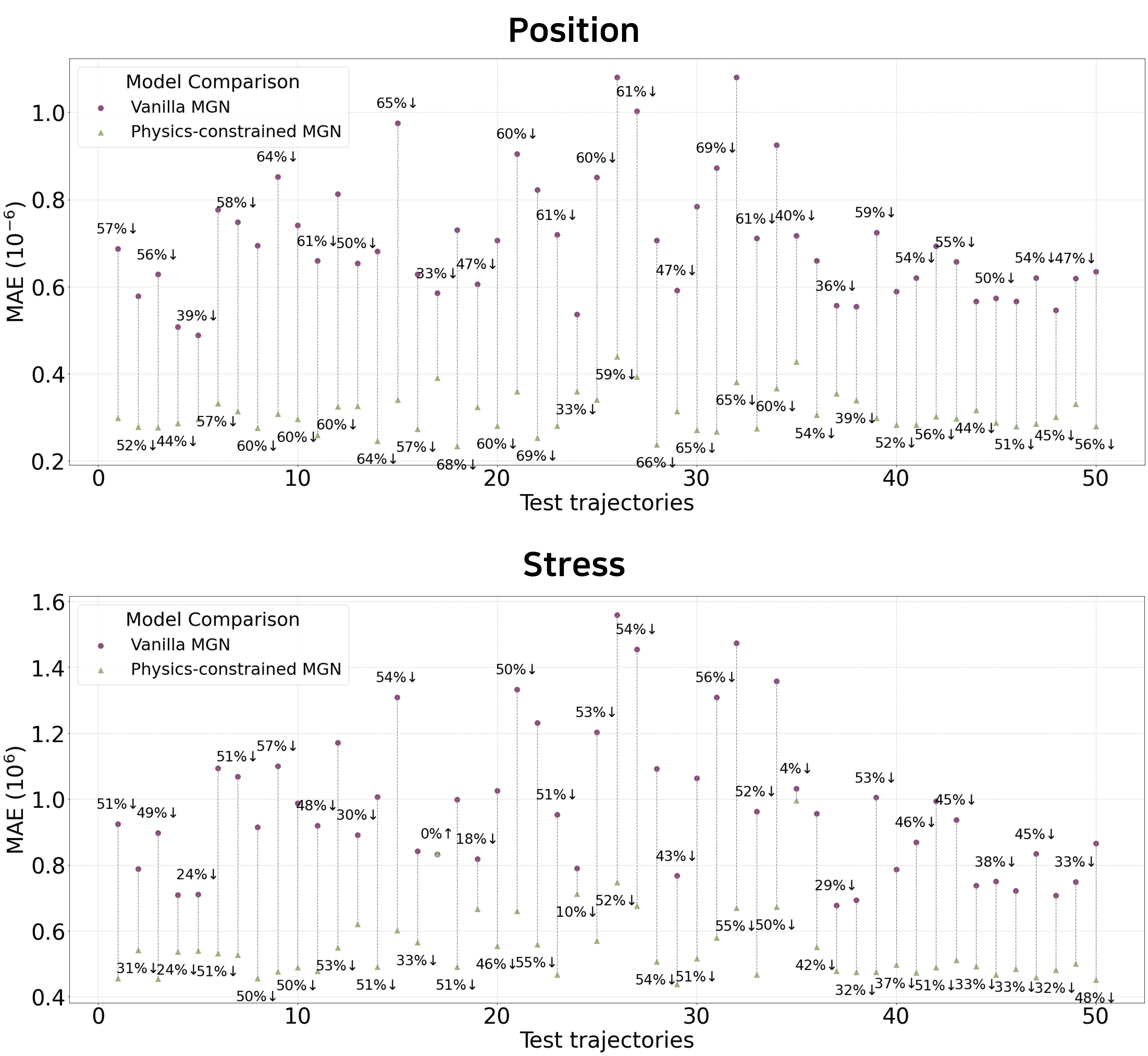}
    \captionsetup{justification=raggedright,singlelinecheck=false}
    \caption{Comparison of prediction performance between vanilla MGN and physics-constrained MGN per test trajectory. Top: Position prediction errors (MAE across rollout-all). Bottom: Stress prediction errors (MAE across rollout-all). Each percentage indicates the relative decrease in physics-constrained MAE compared to vanilla MAE.}
    \label{fig:Fig.13}
\end{figure}

Lastly, the physics-constrained MGN proves that it universally maintains higher performance across all test trajectories, as shown in \cref{fig:Fig.13}. The percentages at each trajectory indicate the relative error reduction achieved by our physics-constrained model (green triangular markers) compared to the vanilla MGN (purple circular markers). This consistent improvement in both position and stress predictions emphasizes the remarkable robustness and reliability of our physics-constrained approach throughout all of the test cases, making it a valuable tool for accurate OLED panel design and optimization across diverse scenarios.

\subsection{Investigation and enhancement of model robustness and generalization capabilities}
\subsubsection{Effects of noise injection: A case study across various noise magnitudes and components}

In this section, we assess the model's robustness with noise injection approaches based on Gaussian white noise. By applying additive white Gaussian noise to the input data during training, the model's adaptability is expected to be enhanced, enabling it to solve a wide range of complex forward and inverse problems \citep{sanchez2020learning}. This technique introduces small perturbations into the input data (in our case, the global position \(x^w\)) used for training, with these perturbations following a Gaussian distribution with a mean of 0 and a standard deviation of \(\sigma\). Acting as a powerful regularization tool, it helps the model relieve the error accumulation, which can cause distribution shifts during long-term rollouts \citep{toshev2023relationships, yang2024enhancing}.

\begin{table}[H]
\centering
\tabularfont
\renewcommand{\arraystretch}{1.3}
\begin{adjustbox}{max width=\textwidth}
\begin{tabular}{c|c|cccccccc}

\hline
\multicolumn{2}{c|}{} & \multicolumn{8}{c}{\textbf{Injected noise magnitudes}} \\
\cline{3-10}
\multicolumn{2}{c|}{} & \multicolumn{2}{c}{0.1$\mu$m} & \multicolumn{2}{c}{\textbf{0.3$\bm{\mu}$m}} & \multicolumn{2}{c}{0.7$\mu$m} & \multicolumn{2}{c}{1.5$\mu$m} \\
\cline{3-10}
\multicolumn{2}{c|}{} & Position & Stress & Position & Stress & Position & Stress & Position & Stress \\
\hline
\multirow{3}{*}{\begin{tabular}[c]{@{}c@{}}Vanilla\\MGN\end{tabular}} & MAE & 3.458 & 2.588 & 0.704 & 0.978 & 0.915 & 0.953 & 1.806 & 1.551 \\
& RMSE & 14.733 & 16.319 & 2.509 & 11.744 & 4.544 & 3.914 & 9.318 & 9.728 \\
& Time [h]& \multicolumn{2}{c}{19.500} & \multicolumn{2}{c}{19.617} & \multicolumn{2}{c}{19.617} & \multicolumn{2}{c}{21.300} \\
\hline
\multirow{3}{*}{\textbf{\begin{tabular}[c]{@{}c@{}}Physics-\\constrained\\MGN\end{tabular}}} & MAE & 1.203 & 1.685 & \textbf{0.309} & \textbf{0.547} & 1.421 & 1.786 & 1.504 & 1.815 \\
& RMSE & 4.308 & 11.766 & \textbf{0.707} & \textbf{2.223} & 2.446 & 7.140 & 4.464 & 11.824 \\
& Time [h] & \multicolumn{2}{c}{21.433} & \multicolumn{2}{c}{21.233} & \multicolumn{2}{c}{22.133} & \multicolumn{2}{c}{22.267} \\
\hline
\end{tabular}
\end{adjustbox}
\captionsetup{justification=raggedright, singlelinecheck=false}
\caption{Investigation of predictive model performance with varying noise magnitudes across all nodes. Note: position errors (MAE, RMSE) are in units of \(10^{-6}\), while stress errors (MAE, RMSE) are in units of \(10^{6}\).}
\label{table:noise_magnitudes}
\end{table}

To find the optimal noise magnitudes, we tested four standard deviation values: 0.1, 0.3, 0.7, and 1.5$\mu$m, considering the scale of the training data. As shown in \Cref{table:noise_magnitudes}, while the training time remained relatively stable across different noise injection scenarios (since noise injection process requires insignificant computational cost), the impact of noise magnitude on model performance was significant. Our proposed model showed the best performance with a noise of 0.3$\mu$m, reducing position error by 71.8\% and stress error by 81.1\% compared to the best vanilla MGN, based on RMSE. Ultimately, this means that by injecting noise levels appropriate to each model, the model can learn to compensate for perturbations that degrade long-term rollout performance through error accumulation.

\begin{table}[H]
\centering
\tabularfont
\renewcommand{\arraystretch}{1.3} 
\begin{adjustbox}{max width=\textwidth}
\begin{tabular}{c|c|cccccccc}

\hline
\multicolumn{2}{c|}{} & \multicolumn{8}{c}{\textbf{Body components for noise injection}} \\
\cline{3-10}
\multicolumn{2}{c|}{} & \multicolumn{2}{c}{None} & \multicolumn{2}{c}{Ball} & \multicolumn{2}{c}{Display panels} & \multicolumn{2}{c}{\textbf{All}} \\
\cline{3-10}
\multicolumn{2}{c|}{} & Position & Stress & Position & Stress & Position & Stress & Position & Stress \\
\hline
\multirow{3}{*}{\begin{tabular}[c]{@{}c@{}}Vanilla\\MGN\end{tabular}} & MAE & 2.536 & 4.431 & 6.456 & 4.119 & 6.685 & 4.397 & 0.704 & 0.978 \\
& RMSE & 10.788 & 26.206 & 32.351 & 23.608 & 32.727 & 26.705 & 2.509 & 11.744 \\
& Time & \multicolumn{2}{c}{19.200} & \multicolumn{2}{c}{19.183} & \multicolumn{2}{c}{20.400} & \multicolumn{2}{c}{19.617} \\
\hline
\multirow{3}{*}{\textbf{\begin{tabular}[c]{@{}c@{}}Physics-\\constrained\\MGN\end{tabular}}} & MAE & 3.826 & 4.465 & 3.020 & 4.747 & 7.166 & 5.380 & \textbf{0.309} & \textbf{0.547} \\
& RMSE & 18.346 & 25.339 & 11.459 & 28.702 & 34.108 & 33.516 & \textbf{0.707} & \textbf{2.223} \\
& Time & \multicolumn{2}{c}{21.283} & \multicolumn{2}{c}{21.283} & \multicolumn{2}{c}{21.367} & \multicolumn{2}{c}{21.233} \\
\hline
\end{tabular}
\end{adjustbox}
\captionsetup{justification=raggedright, singlelinecheck=false}
\caption{Investigation of predictive model performance with noise injection in each flexible body component. Note: position errors (MAE, RMSE) are in units of \(10^{-6}\), while stress errors (MAE, RMSE) are in units of \(10^{6}\).}
\label{table:noise_injection_components}
\end{table}

We also conducted additional experiments to verify the effects of noise when injected into different flexible body components. We consider four cases: None, Ball, Display panels, and All. \Cref{table:noise_injection_components} presents the experimental results of the two comparative models based on noise injection for each component case. Both models achieved peak performance when noise was introduced into all system components. This noise-induced enhancement suggests that small perturbations to all flexible body components in impact scenarios can effectively mimic the inherent uncertainties and fluctuations present in object interactions, and therefore reduce the error exacerbated through rollout processes.

\begin{figure}[htb!]
    \centering
    \includegraphics[width=0.7\linewidth]{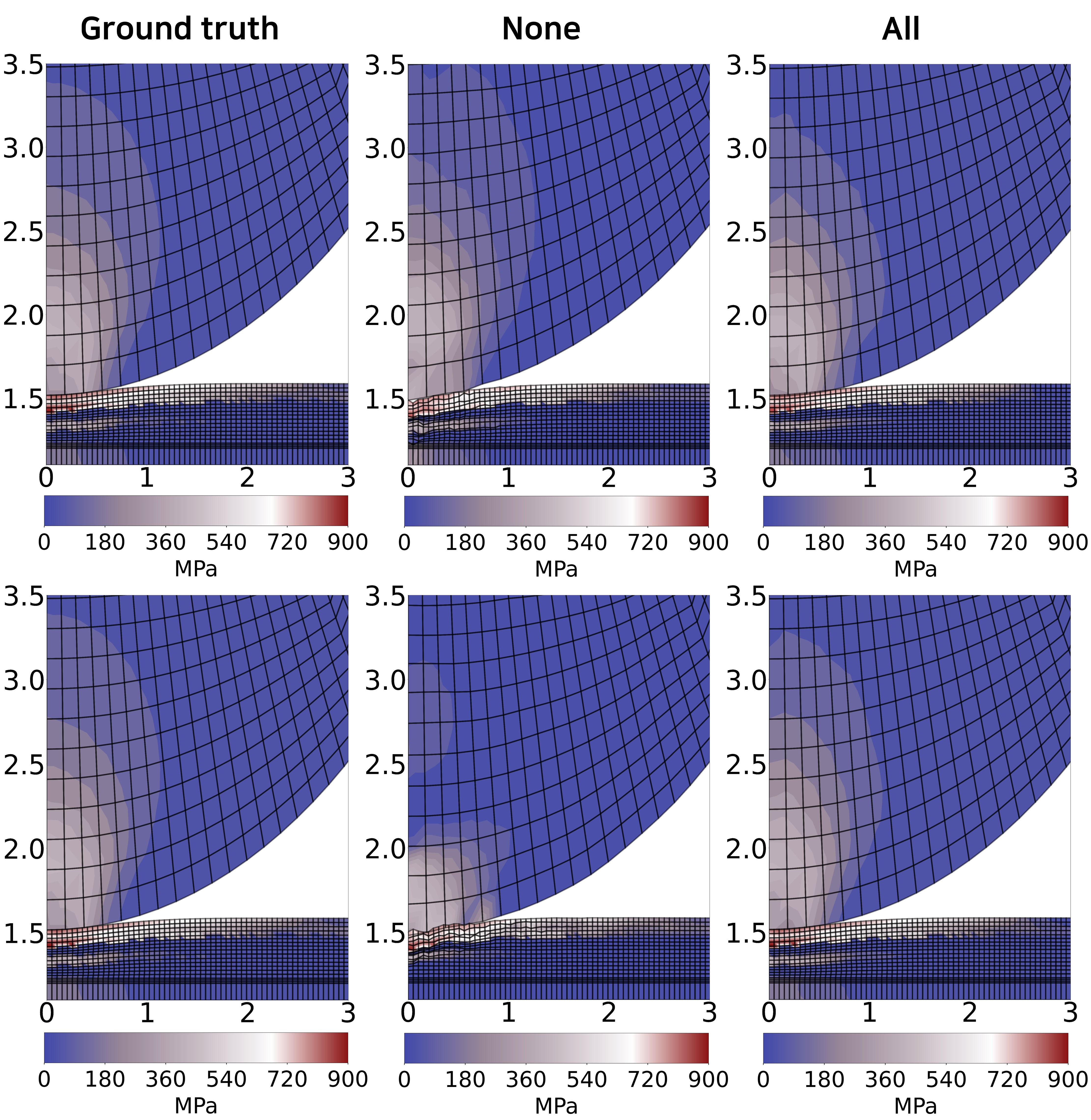}
    \captionsetup{justification=raggedright,singlelinecheck=false}
    \caption{Stress distribution when maximum stress occurs, showing the effect of noise injection into flexible bodies. Top row: vanilla MGN; Bottom row: physics-constrained MGN. Columns represent ground truth (left), without noise injection (middle), and noise injection into all components (right).}
    \label{fig:Fig.14}
\end{figure}

Finally, to provide an intuitive comparison of prediction performance according to the noise injection, we visualized the predicted behavior and stress distribution when the maximum stress occurs during the inference stage (\cref{fig:Fig.14}). It indicates that without the addition of small perturbations, the model's accuracy is generally lower. In the absence of noise (None case), the model struggles to accurately capture the complex interactions between two objects, particularly through contact nodes and edges: they show unrealistic behavior, such as zigzagging nodes in the upper plate. In contrast, when small perturbations are added to all objects (All case), we observe that the model successfully handles the interactions, producing results very similar to the ground truth for both vanilla and physics-constrained models. Through this, we prove the effectiveness of noise injection in improving the model's ability to generalize and accurately predict complex object interactions in terms of rollout process.

\subsubsection{Extrapolation performance: thickness-based predictions beyond training range}

\begin{figure}[htb!]
    \centering
    \includegraphics[width=0.9\linewidth]{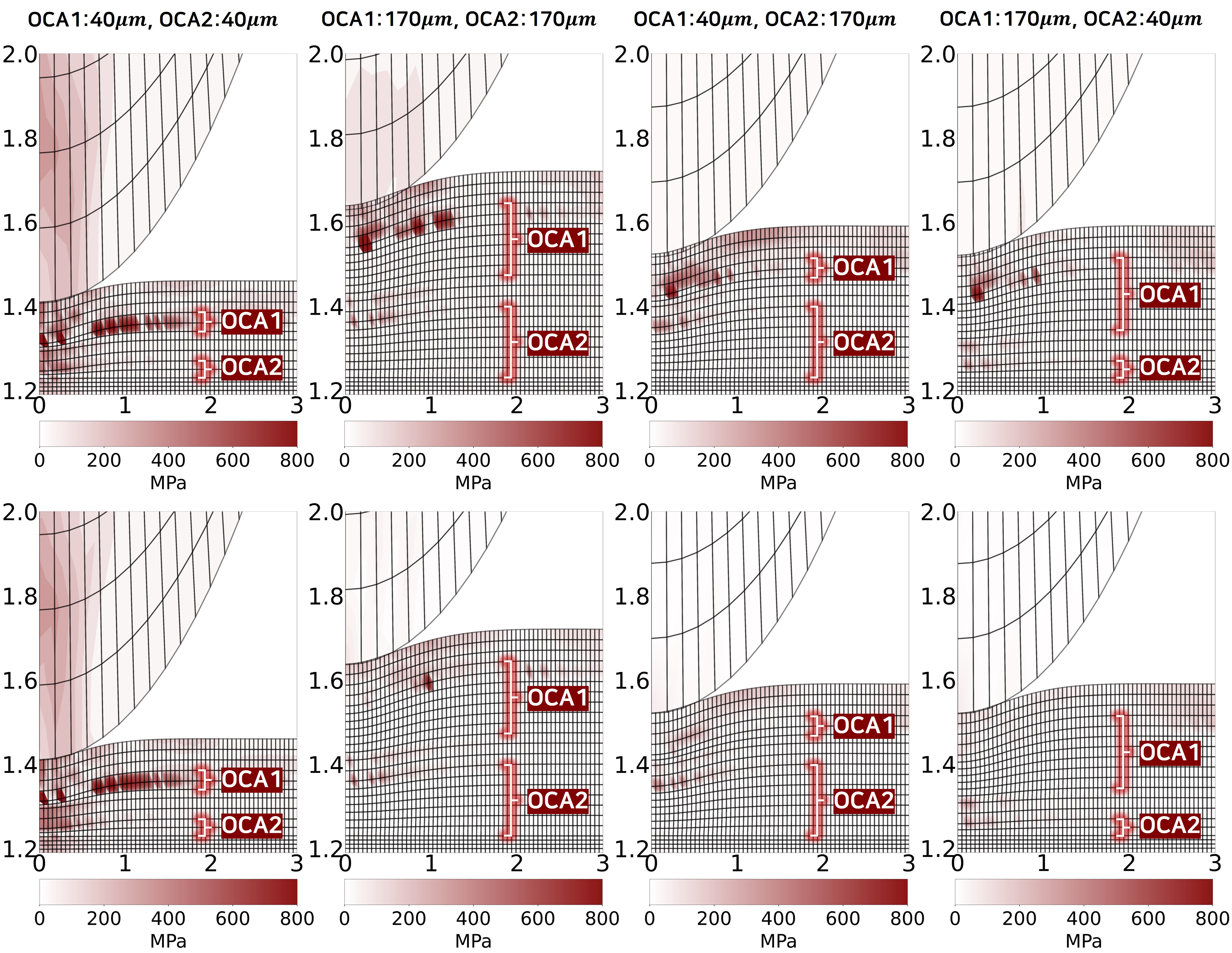}
    \captionsetup{justification=raggedright,singlelinecheck=false}
    \caption{Stress error distribution for extrapolation cases across OCA thicknesses outside the training range. Top row: vanilla MGN; Bottom row: physics-constrained MGN. $\text{OCA}_1$ and $\text{OCA}_2$ values show thicknesses (in $\mu$m) for each extrapolation case. The time steps at which the maximum stress occurs in each case are 30, 49, 43, 43.}
    \label{fig:Fig.15}
\end{figure}

In this section, we evaluate the extrapolation capabilities of the models outside the training range with respect to thickness conditions to validate their practical applicability and overall robustness. As shown in \Cref{table:material_properties}, the OCA thickness in our training dataset ranged from 50 to 150$\mu$m. To assess the models' extrapolation capabilities, we conduct tests under four additional thickness conditions outside this range (\cref{fig:Fig.15}). We examine the time-step when maximum stress occurs for each extrapolation thickness condition. When $\text{OCA}_1$ and $\text{OCA}_2$ both have a thickness of 40$\mu$m (first column in \cref{fig:Fig.15}), where the thickness of both OCAs is less than the training region so this can be considered the most extreme extrapolation case \footnote{Extrapolating to thinner plates is more challenging than extrapolating to thicker plates due to increased sensitivity to small changes in geometry. Thinner plates experience higher stress concentrations, which magnifies the influence of any inaccuracies in the model's learned behavior. Thicker plates, by contrast, tend to distribute stresses more uniformly, making predictions less sensitive to the extrapolated conditions.}, both models have the largest stress error in the $\text{OCA}_1$ plate. This indicates difficulty in accurately predicting complex stress distributions in thin OCA structures. The remarkable decrease in accuracy for thicknesses below the training range suggests limited capability in capturing nonlinear behavior under unseen extreme conditions, indicating that additional approaches or incorporation of training data are required for predictions in these areas. For other thickness conditions (second to fourth column in \cref{fig:Fig.15}), vanilla MGN continues to display penetration phenomena and higher errors, particularly in the CG and $\text{OCA}_1$ regions. In contrast, physics-constrained MGN generally avoids penetration and maintains significantly lower stress errors. This demonstrates that incorporating physical laws into the learning process enhances the model's adaptability across a extrapolation conditions.

\subsubsection{Performance validation of rollout predictions based on output quantities}

As described in Eq. \ref{eq:model inference}, rollout prediction is performed using the forward Euler method. In this section, we evaluate the performance of rollout with different output quantities.

First, we directly predict the absolute node position for the subsequent time step, which simplifies the procedure by eliminating the need for intermediate calculations (Eq. \ref{eq:position}). Second, we predict relative changes, specifically displacement, which enhances the model's sensitivity to capture minor variations effectively (Eq. \ref{eq:displacement}). Finally, we predict velocity, and then use the forward Euler method to estimate the node position at the next time step.

\begin{equation} x_i(t + 1) = \hat{x}_i(t + 1) 
\label{eq:position} 
\end{equation}
\begin{equation} x_i(t + 1) = x_i(t) + \Delta \hat{x}_i 
\label{eq:displacement} 
\end{equation}
 
Here, \(x_i(t)\) represents the position of node \(i\) at time \(t\), \(\Delta \hat{x}_i\) represents the predicted displacement, capturing the difference between the next and current node positions.

\begin{table}[H]
\centering
\tabularfont
\renewcommand{\arraystretch}{1.3}
\begin{adjustbox}{max width=\textwidth}
\begin{tabular}{c|c|cccccc}
\hline
\multicolumn{2}{c|}{} & \multicolumn{2}{c}{Position} & \multicolumn{2}{c}{Displacement} & \multicolumn{2}{c}{Velocity} \\
\cline{3-8}
\multicolumn{2}{c|}{} & \multicolumn{1}{c}{Position} & \multicolumn{1}{c}{Stress} & \multicolumn{1}{c}{Position} & \multicolumn{1}{c}{Stress} & \multicolumn{1}{c}{Position} & \multicolumn{1}{c}{Stress} \\
\hline
\multirow{3}{*}{\begin{tabular}[c]{@{}c@{}}Vanilla\\MGN\end{tabular}} & MAE & 1458.494 & 59.638 & 1.051 & 1.130 & 0.704 & 0.978 \\
& RMSE & 2218.880 & 85.306 & 3.148 & 7.718 & 2.509 & 11.744 \\
& Time [h]& \multicolumn{2}{c}{19.500} & \multicolumn{2}{c}{19.433} & \multicolumn{2}{c}{19.617} \\
\hline
\multirow{3}{*}{\textbf{\begin{tabular}[c]{@{}c@{}}Physics-\\constrained\\MGN\end{tabular}}} & MAE & 1318.342 & 34.451 & 0.485 & 0.797 & \textbf{0.309} & \textbf{0.547} \\
& RMSE & 2070.163 & 60.919 & 1.268 & 3.472 & \textbf{0.707} & \textbf{2.223} \\
& Time [h]& \multicolumn{2}{c}{21.383} & \multicolumn{2}{c}{21.833} & \multicolumn{2}{c}{21.233} \\
\hline
\end{tabular}
\end{adjustbox}
\captionsetup{justification=raggedright, singlelinecheck=false}
\caption{Comparison of rollout prediction performance between the vanilla MGN model and the physics-constrained MGN model based on three output quantities.}
\label{table:rollout predictions}
\end{table}

As shown in \Cref{table:rollout predictions}, our experiments indicate that direct position prediction leads to significant prediction errors. In contrast, the velocity prediction using the forward Euler method combined with the physics-constrained MGN proves to be the most effective. This approach updates the velocity at each time step and utilizes it to predict the position at the next time step (Eq. \ref{eq:model inference}), enabling the model to accurately capture continuous motion and provide highly precise predictions for both position and stress, including the fine-scale variations inherent in Lagrangian systems. Additionally, displacement-based predictions show significantly lower errors compared to directly predicting absolute node positions at the next time step, though they still fall short of velocity-based predictions. The superior performance of both velocity-based and displacement-based predictions over direct absolute position predictions highlights the advantages of using relative changes to estimate the next node position.

\section{Extension to design optimization}
\label{sec:Sec5}

\subsection{Problem definition for design optimization of OCA thickness}

This section focuses on using the trained MGN models for the design optimization of OCA thickness. To determine the optimal configuration of multi-layered display panels that minimizes stress distribution in the OLED panel, we optimize the thickness of the two OCA layers using a genetic algorithm (GA) with a single objective. As outlined in Eq. \ref{eq:optimization}, the optimization problem is formulated with specific constraints:

\begin{equation}
\begin{aligned}
    &\min_{\mathbf{X}} \quad \hat{f}(\mathbf{X}) \\
    &\text{w.r.t} \quad \mathbf{X} = \begin{bmatrix} T_{\text{OCA}_1}, T_{\text{OCA}_2} \end{bmatrix} \\
    &\text{subject to} \quad 50\text{$\mu$m} \leq \mathbf{X} \leq 150\text{$\mu$m}, \\
    &\quad 150\,\text{$\mu$m} \leq T_{\text{OCA}_1} + T_{\text{OCA}_2} \leq 250\,\text{$\mu$m}, \\
\end{aligned}
\label{eq:optimization}
\end{equation}

Here, $\hat{f}(\mathbf{X})$ is calculated as the average stress of nodes where $x < 3$ in the mesh grid: this region is specifically chosen because it represents the area where intensive stress occurs due to impact, making it crucial for optimizing stress distribution. 

The optimization process, using surrogate models to accelerate computations, was completed in 2 hour and 24 minutes. We employed a genetic algorithm (GA) with a population size of 100, running for 50 generations, to minimize stress distribution. Given the nature of spatio-temporal prediction, most of the computation time was spent on performing rollouts, which scales with the population and generation size. However, once the initial shape is established, full simulations for various scenarios can be efficiently validated, making this approach significantly more efficient compared to direct numerical simulation.

\subsection{Comparison of consistency between ground truth and model predictions}

In this section, we compare the consistency between ground truth data and the predictions made by both the vanilla MGN and the physics-constrained MGN, focusing on the acquired optimal solution. The purpose of this comparison is to verify how accurately each model predicts the stress distribution in OLED panels and thus how much validity the optimal solution has compared to the ground truth solver, which in our case is an explicit dynamics solver. Assessing this consistency is crucial for validating the reliability of each model and highlights the significance of achieving accurate optimal results in real-time through surrogate-based stress prediction.

\begin{table}[H]
\centering
\renewcommand{\arraystretch}{1.3}
\begin{adjustbox}{max width=\textwidth}
\begin{tabular}{c|cc}
\hline
 & \multicolumn{2}{c}{OCA Thickness Optimization} \\
\cline{2-3}
 & Vanilla MGN & Physics-constrained MGN \\
\hline
Optimal values ($\mu$m) & 
\begin{tabular}[c]{@{}c@{}}
$T_{\text{OCA}_1}$ \(=100.364\), \\
$T_{\text{OCA}_2}$ \(=149.635\) 
\end{tabular} & 
\begin{tabular}[c]{@{}c@{}}
$T_{\text{OCA}_1}$ \(=145.207\), \\
$T_{\text{OCA}_2}$ \(=99.322\) 
\end{tabular} \\
\hline
Mean stress of ground truth & 1.995 MPa & 1.891 MPa \\
\hline
Mean stress of prediction & 1.666 MPa & 1.860 MPa \\
\hline
Error & 16.45\% & \textbf{1.67\%} \\
\hline
\end{tabular}
\end{adjustbox}
\caption{Comparison of optimization results for vanilla and physics-constrained models.}
\label{table:comparison optimization results}
\end{table}

\Cref{table:comparison optimization results} summarizes the optimal OCA layer thickness values derived from each predictive model, the mean stress values within OLED panels from both ground truth solver and model predictions, and the percentage error between those stress values. The noteworthy point is that those two MGN models derived two optimal solutions with totally different OCA thickness: vanilla gets optimal design with maximum $\text{OCA}_2$ value whereas physics-constrained model obtains design with maximum $\text{OCA}_1$ value. While the vanilla MGN initially seemed promising with lower stress values, comparison with the ground truth revealed that the physics-constrained MGN produced a significantly more accurate result. Evaluating these optimal designs, the results indicate that the physics-constrained MGN demonstrates superior consistency, reducing the error rate by approximately 89.8\%, compared to the vanilla MGN model. Our proposed model significantly improves accuracy in predicting actual stress distribution, thereby providing a more reliable method for optimizing OLED panel design.

\section{Conclusions and future work}
\label{sec:Sec6}

While conventional MGNs have shown promise in spatio-temporal prediction tasks, their exclusive reliance on data-driven modeling presents significant challenges in accurately representing physical phenomena such as drop impact tests. In this study, we proposed a physics-constrained MGN to address the challenges of accurately predicting the nonlinear dynamics of multi-layered display panels under ball drop impact. Specifically, our model incorporates physical constraints directly into the existing loss function by calculating the penetration depth between objects and adding it as a weighted loss term. This approach ensures compliance with the fundamental physical principle of non-penetration during object contact. To demonstrate the effectiveness of our proposed model, we first conducted parametric study demonstrating the importance of polynomial degree selection for object boundary approximations and then explored the effects of the penetration loss weight ($\lambda$) that balances accuracy and stability in drop impact scenario. Results indicated that our physics-constrained MGN significantly outperformed the vanilla MGN with reduced error accumulation during rollout predictions owing to its physics-constrained loss function. Moreover, it universally achieved lower errors across all test datasets, effectively mitigating penetration issues also in the unseen impact cases. We additionally validated the model's robustness through various noise injection experiments, adjusting both noise magnitude and target objects to be contaminated by the noise. Introduction of noise across both ball and plate resulted in significantly lower errors than the case without noise, while maintaining consistent spatio-temporal accuracy at each time step, even with previously unseen data. Furthermore, we also demonstratedthe effects of rollout predictions based on different physical output quantities, showing that predicting relative changes (such as displacement and velocity) is significantly more stable than directly predicting absolute positions at the next time step. Then, we confirmed the model's extrapolation performance by exploring beyond the training range of design variables. Except for the excessive condition, where the both design variables were thinner than the training range, the physics-constrained model showed the superior extrapolation performance compared to the baseline vanilla model. Finally, we applied our physics-constrained MGN as a spatio-temporal surrogate model in the design optimization procedure for optimizing the thickness of OCA layers to minimize stress distribution in OLED panels. The optimal solution obtained by the proposed model shows better performance than that obtained by the vanilla model, and also the discrepancy between the ground truth (computer solver) and the prediction by the models is found to be much smaller in the physically constrained model compared to the vanilla model.

Here are the primary outcomes of our research:
\begin{enumerate}[itemsep=10pt]

\item The results demonstrated that selecting the appropriate combination of polynomial degrees for modeling of each object boundary and introducing a penetration loss weight of $\lambda = 10$ led to significant improvements in model performance. Compared to the model without applying polynomial approximations and a penetration loss weight, our model achieved a 71.8\% decrease in position RMSE and an 81.1\% decrease in stress RMSE. 

\item Injecting noise into both ball and plate with a magnitude of 0.3 $\mu$m yielded the best results, reducing errors by 64.3\% in position MAE and 81.1\% in stress MAE compared to the case without noise injection. 

\item We further found that the rollout with velocity-based prediction achieved a 70.5\% reduction in position MAE and a 51.6\% reduction in stress MAE compared to direct position prediction. In this regard, we demonstrated that predicting relative changes, such as displacement or velocity, captures local variations and dynamics, which helps reduce cumulative errors during rollouts and is significantly more stable and effective than directly predicting absolute positions.

\item We applied the physics-constrained MGN for the design optimization of OCA layer thickness, achieving an optimal solution with an error rate of 1.67\% compared to the ground truth, which represents an approximate 89.8\% reduction in error compared to the vanilla MGN's optimal solution. It demonstrates that the proposed model provides not only a more accurate prediction but also better optimization outcomes for minimizing stress distribution in OLED panels.

\end{enumerate}

As part of our future work, we aim to enhance the integration of physical laws within our modeling framework. Currently, our approach indirectly incorporates physics constraint through purely data-driven approach; however, we plan to investigate more direct methods of integrating physics, such as physics-informed neural networks \cite{yang2024data}, to achieve an analytic representation of physical laws and improve prediction accuracy further. In addition, we seek to reduce training time through graph pooling techniques and investigate the use of hypergraph structures that account for relationships between nodes and elements, rather than only considering node-edge interactions as in our current graph structure. This approach aims to strike a balance between accuracy and computational efficiency, ultimately enhancing the overall performance of our models.

\section*{CRediT authorship contribution statement}
\textbf{Jiyong Kim:} Conceptualization, Formal analysis, Investigation, Methodology, Visualization, Writing - Original draft, Writing – Review \& Editing, Supervision, Software. \textbf{Jangseop Park:} Methodology, Software. \textbf{Sunwoong Yang:} (Co-corresponding authors) Methodology, Supervision, Writing – Review \& Editing. \textbf{Namwoo Kang:} (Co-corresponding authors), Project administration, Resources, Supervision.

\section*{Declaration of competing interest}
The authors declare that they have no known competing financial interests or personal relationships that could have ap peared to influence the work reported in this paper.

\section*{Data availability}
Data will be made available on request.

\section*{Acknowledgements}
This work was supported by LG display, the National Research Foundation of Korea grant (2018R1A5A7025409), the Ministry of Science and ICT of Korea grant (No. 2022-0-00969, No. 2022-0-00986, and No. RS-2024-00355857), the Ministry of Trade, Industry \& Energy grant (RS-2024-00410810), and the National Research Council of Science \& Technology (NST) grant by the Korea government (MSIT) (No. GTL24031-300).

\makeatletter
\renewcommand\@biblabel[1]{}
\makeatother

\bibliographystyle{elsarticle-harv}

\end{document}